\documentclass[11pt]{article}
\usepackage{subfigure}
\usepackage{graphics}
\usepackage{amssymb}
\usepackage{epsf}
\usepackage{epsfig}
\usepackage{rotating}
\usepackage{pstricks}
\usepackage{color}
\usepackage{url} 

\usepackage{cite} 
\usepackage[margin=2.cm]{geometry}
\def\ds#1{#1\kern-1ex\hbox{/}}
\def\dsh{h\kern-1.2ex /}

\newcommand{\bea}{\begin{eqnarray}}
\newcommand{\eea}{\end{eqnarray}}

\def\beq{\begin{equation}}
\def\eeq{\end{equation}}

\def\beqn{\begin{eqnarray}}
\def\eeqn{\end{eqnarray}}
\def\ba{\begin{eqnarray}}
\def\ea{\end{eqnarray}}

\newcommand{\be}{\begin{equation}}
\newcommand{\beqa}{\begin{eqnarray}}
\newcommand{\eeqa}{\end{eqnarray}}

\newcommand{\ee}{\end{equation}}
\begin{document}
\noindent
DESY 14-077\\
LPN 14-076\\
SFB/CPP-14-27\\
\hfill
\begin{center}

\vspace{1.0cm}
{\bf \large Top-quark pair production at hadron colliders: 
differential cross section and phenomenological applications with DiffTop}
\vspace{0.5cm}

\vspace{1cm}
{ \bf Marco Guzzi$^a$, Katerina Lipka$^a$ and Sven-Olaf Moch$^{b,c}$}

%\vspace{1cm}

{\it $^a$Deutsches Elektronen-Synchrotron DESY, \\Notkestrasse 85, D-22607 Hamburg, Germany\\}
{\it $^b$II.~Institut f\"ur Theoretische Physik, Universit\"at Hamburg, \\Luruper Chaussee 149, D-22761 Hamburg, Germany\\}
{\it $^c$Deutsches Elektronen-Synchrotron DESY,\\ Platanenallee 6, D-15738 Zeuthen, Germany}
\vspace{.5cm}\footnote{marco.guzzi@desy.de, katerina.lipka@desy.de, sven-olaf.moch@desy.de}

\begin{abstract}

The results of phenomenological studies of top-quark pair production
in proton-proton collisions are presented. Differential cross sections are calculated in perturbative 
QCD at approximate next-to-next-to-leading order ${\cal O}(\alpha_s^4)$ by using 
methods of threshold resummation beyond the leading logarithmic accuracy. Predictions for 
the single-particle inclusive kinematics are presented for transverse momentum and rapidity distributions 
of final-state top quarks. Uncertainties related to the description of proton structure, top-quark mass and 
strong coupling constant are investigated in detail.
The results are compared to the recent measurements by the ATLAS and CMS collaborations 
at the LHC at the center of mass energy of 7 TeV. The calculation presented here is implemented 
in the computer code \textsc{Difftop} and can be applied to the general case of heavy-quark pair 
production at hadron-hadron colliders. For the first time, a fit of parton distribution functions at NNLO
is performed by using the differential cross sections of top-quark pair production together with other data sets. 
The impact of the top-pair production on the precision of the gluon distribution at high scales is illustrated.

\end{abstract}
\end{center}
\date{\today}
\newpage

\section{Introduction}

Studies of heavy-quark production at hadron colliders provide stringent tests of quantum chromodynamics 
(QCD) and of the theory of electroweak (EW) interactions. Furthermore, these are of crucial 
importance in searches for signatures of physics Beyond the Standard Model (BSM). The mass of the recently discovered Higgs 
boson~\cite{Aad:2012tfa,Chatrchyan:2012ufa} has been measured to be in the range $120 \leq m_H \leq 135$ GeV, 
therefore the Higgs sector is expected to be closely related to the physics of the top-quark. In particular, 
the role of quantum corrections to the top-quark mass, which together with the mass of the Higgs-boson define
the electroweak vacuum stability conditions has been studied~\cite{Alekhin:2012py,Bezrukov:2012sa}.

Experimentally, top-quark physics is being studied at the Tevatron and is extensively 
explored at the unprecedented energies of the Large Hadron Collider (LHC). The CMS, ATLAS, D\O, and 
CDF collaborations have recently combined their results~\cite{ATLAS:2014wva} for the top-quark mass 
and obtained a value of $m_t=173.3\pm 0.76$ GeV. The interpretation of the measured mass and its relation to the 
theoretically well-defined pole mass of the top quark is discussed in details in~\cite{Moch:2014tta}. New 
observables to be used for the measurement of the top-quark mass at hadron colliders are 
proposed in~\cite{Alioli:2013mxa, Biswas:2010sa}. The most recent determination of the pole mass of 
the top quark is performed by the CMS collaboration~\cite{Chatrchyan:2013haa}. In the same analysis, 
for the first time, the issue of correlations between the top-quark pole mass, gluon distribution, and strong 
coupling constant $\alpha_s$ in QCD predictions for the inclusive cross section of top-quark pair 
production is discussed.

Precise measurements for the total and differential cross section for top-quark pair production at a 
center-of-mass energies $\sqrt{S}=7$ and 8 TeV have been recently published by the CMS~\cite{Chatrchyan:2012saa,Chatrchyan:2013faa} and ATLAS~\cite{Aad:2012hg,Aad:2014zka,TheATLAScollaboration:2013eja,TheATLAScollaboration:2013dja} collaborations. 
The interpretation of current and forthcoming LHC data demands high-precision theory predictions 
that imply a new realm of precision calculations in perturbative QCD (pQCD), supplied by the development 
of efficient tools for phenomenological analyses. The QCD corrections to heavy-quark production at hadron colliders at the next-to-leading order 
(NLO), ${\cal O}(\alpha_s^3)$, are known since many years~\cite{Nason:1987xz,Nason:1989zy,Beenakker:1988bq,Meng:1989rp,Beenakker:1990maa,Mangano:1991jk}. The full calculation at next-to-next-to-leading order (NNLO), ${\cal O}(\alpha_s^4)$, 
for the inclusive cross section has been accomplished only recently~\cite{Czakon:2013goa,Czakon:2012pz,Czakon:2012zr,Baernreuther:2012ws} and required continuous efforts of the QCD community in calculating radiative corrections 
and in the development of computational tools~\cite{Czakon:2007wk,Czakon:2007ej,Mitov:2006xs,Ferroglia:2009ep,Ferroglia:2009ii,Czakon:2010td,Bierenbaum:2011gg,Baernreuther:2013caa}. The NNLO calculation of the inclusive cross section for the $t\bar{t}$ production is implemented 
in the \textsc{C++} computer programs \textsc{Top++}~\cite{Czakon:2011xx} and \textsc{Hathor}~\cite{Aliev:2010zk}.
In these calculations the final-state top quarks are considered in the on-shell approximation.
Studies in QCD at NLO where final-state top quarks decay into pairs of $W$ bosons and 
$b$ quarks can be found in Ref.~\cite{Denner:2010jp,Bevilacqua:2010qb,Denner:2012yc}.

The comparison of QCD predictions with the data and a multitude of phenomenological analyses of interest  
require to have precise predictions not only for the inclusive cross section, but also at differential level. 
Invariant mass distribution of $t\bar{t}$, transverse momentum and rapidity distributions of the top quark or 
 $t\bar{t}$, are examples of differential distributions needed at the highest perturbative order possible. 
The exact NLO calculations for $t\bar{t}$ total and differential cross sections are implemented 
into Monte Carlo (MC) numerical codes \textsc{MCFM}~\cite{Campbell:2000bg},
\textsc{MC@NLO}~\cite{Frixione:2003ei}, \textsc{POWHEG}~\cite{Alioli:2010xa}, 
\textsc{MadGraph/MadEvent}~\cite{Alwall:2007st,Frederix:2009yq}. The NNLO corrections for these 
observables are not yet available.

In this paper, we present the results of a phenomenological analysis of differential cross section 
of $t\bar{t}$ production at the LHC, in which the estimate of the uncertainties due to 
the knowledge of the proton structure is addressed.
The recent inclusive and, for the first time, differential measurements
of $t\bar{t}$ production at hadron colliders are included in a QCD analysis at NNLO. 
The impact of such data on the uncertainty of the gluon distribution is illustrated.
For this purpose, the necessary developments of computing tools based on the available  
theory had to be performed. The approximate NNLO ${\cal O}(\alpha_s^4)$ calculation for the 
differential cross section in the single-particle inclusive (1PI) kinematic for heavy-flavor 
production at hadron colliders has been implemented in a novel 
computer code $\textsc{DiffTop}$ which we provide at~\cite{difftop_web}.
In this calculation, techniques of logarithmic expansion beyond the leading 
logarithmic accuracy in QCD threshold resummation are used.

\subsection{Probing the proton structure through $t\bar{t}$ production}

Top-pair production at the LHC probes parton distribution functions (PDFs) of the proton, in 
particular the gluon distribution. In proton-proton collisions at the LHC, $t\bar{t}$ production 
is mainly driven by the gluon, in fact approximately 85\% of the total cross section is ascribed to the gluon-gluon 
channel at $\sqrt{S}=7$ TeV and this fraction grows with the increase of $\sqrt{S}$. Measurements 
of the $t\bar{t}$ total and differential cross sections offer the possibility of probing the 
gluon in the large Bjorken $x$ region ($x\approx 0.1$) where gluon is currently poorly constrained. 
This has first been studied in~\cite{Beneke:2012wb} (see also~\cite{Czakon:2013tha}).
However, strong correlations between $\alpha_s$, gluon distribution $g(x)$, and the 
pole mass of the top quark $m_t$, in the QCD description of $t\bar{t}$ production have to be taken into 
account. A simultaneous determination of $g(x)$, $\alpha_s$ and $m_t$, using the $t\bar{t}$ measurements 
at the LHC in a QCD analysis, together with relevant measurements in Deep-Inelastic Scattering (DIS) 
and in proton-(anti)proton collisions might resolve these correlations. 

The developed program and computational tools used for the phenomenological analyses presented in this paper, 
provide a basis for the inclusion of differential 
$t\bar{t}$ cross sections into QCD analyses at NNLO. For the purpose of a fast 
calculation within QCD analyses for PDF determination, $\textsc{DiffTop}$ is interfaced to 
\textsc{fastNLO}~\cite{dis2014Fast,Britzger:2012bs,Wobisch:2011ij,Kluge:2006xs} and included into 
the QCD analysis platform \textsc{HERAFitter}~\cite{herafitter}. 
This allows the user to perform, for the first time, full genuine PDF fits including differential cross sections
of $t\bar{t}$ production by using fast theoretical predictions for these observables.
This not only represents a clear advantage in terms of the CPU-time consumption,
but also opens the window to fully explore the potential of the $t\bar{t}$ measurements.
The exact NNLO calculation for the fully differential $t\bar{t}$ distributions, once available, 
will be of clear advantage. 
However, full global QCD analyses of the current and forthcoming 
high-precision LHC measurements set specific requirements to the representation 
of the experimental data and availability of fast computing tools.
The present analysis addresses these requirements in the context of differential 
$t\bar{t}$ production cross sections for the first time, by using approximate calculations in order 
to facilitate future PDF fits using the exact NNLO theory.
 
Fast theoretical predictions for the full NNLO differential cross section calculation can be in principle 
obtained by using tools like 
\textsc{fastNLO}~\cite{dis2014Fast,Britzger:2012bs,Wobisch:2011ij,Kluge:2006xs} or 
\textsc{APPLgrid}~\cite{Carli:2010rw}. 
The grid files generated in the output of such codes can be directly 
read inside the \textsc{HERAFitter} open-source platform to determine the impact of the experimental $t\bar{t}$ data 
on PDFs by performing the QCD analysis.  
Presumably, the full NNLO calculation will be lengthy and CPU time consuming 
and the complication for creating an interface to \textsc{fastNLO} or \textsc{APPLgrid} 
will depend upon the structure of the calculation itself and the way the integration is done. 
One can forsee that the generation of \textsc{fastNLO}/\textsc{APPLgrid} grids will be feasible, 
but it might happen that it is dramatically CPU time consuming.
If that will be the case, a possible way out will be to work on approximate calculations 
which can be highly improved once the full calculation is available.

\subsection{Threshold resummation techniques}

\begin{figure}[p]
\begin{center}
\includegraphics[width=7cm, angle=0]{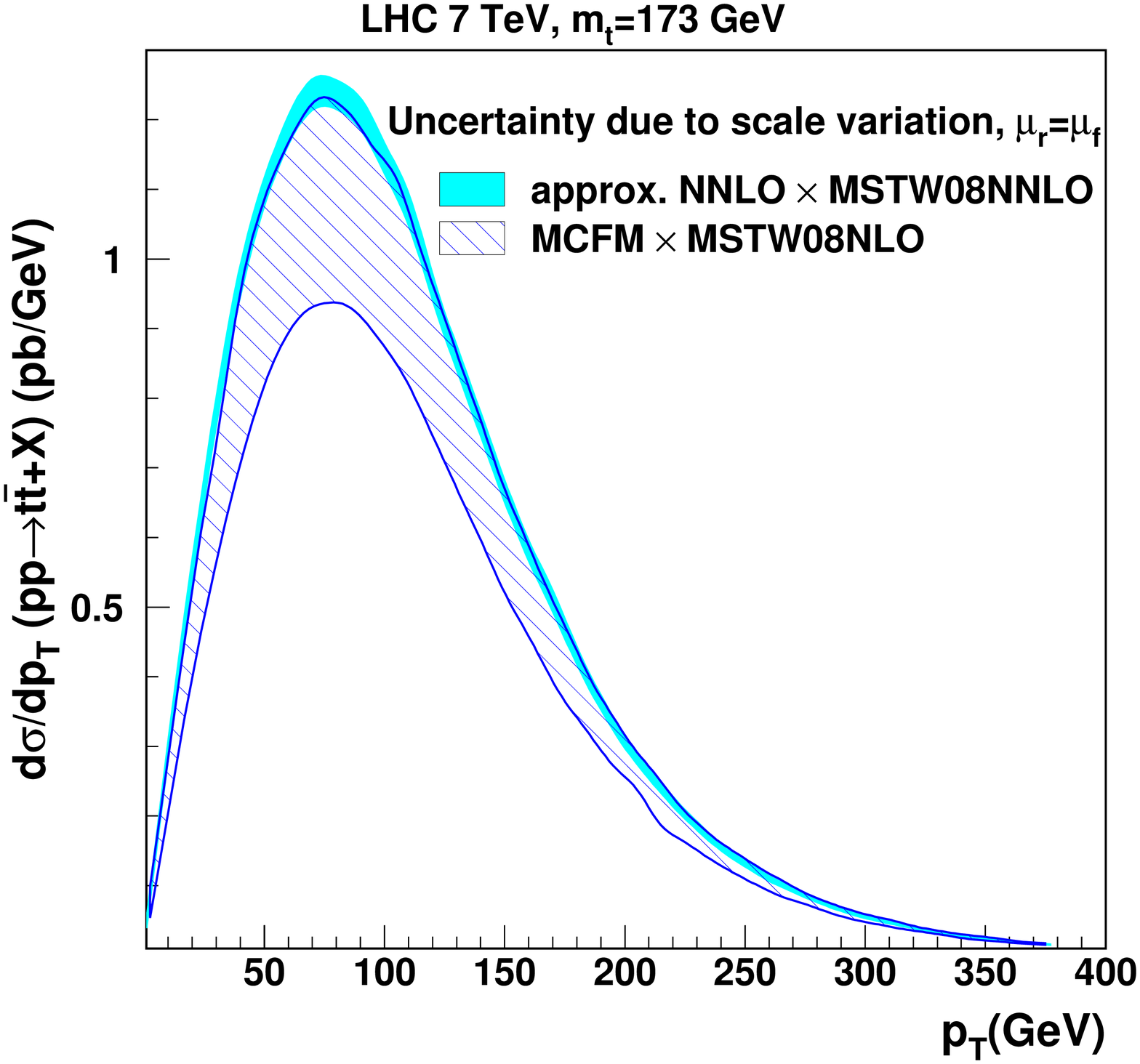}\\
\includegraphics[width=7cm, angle=0]{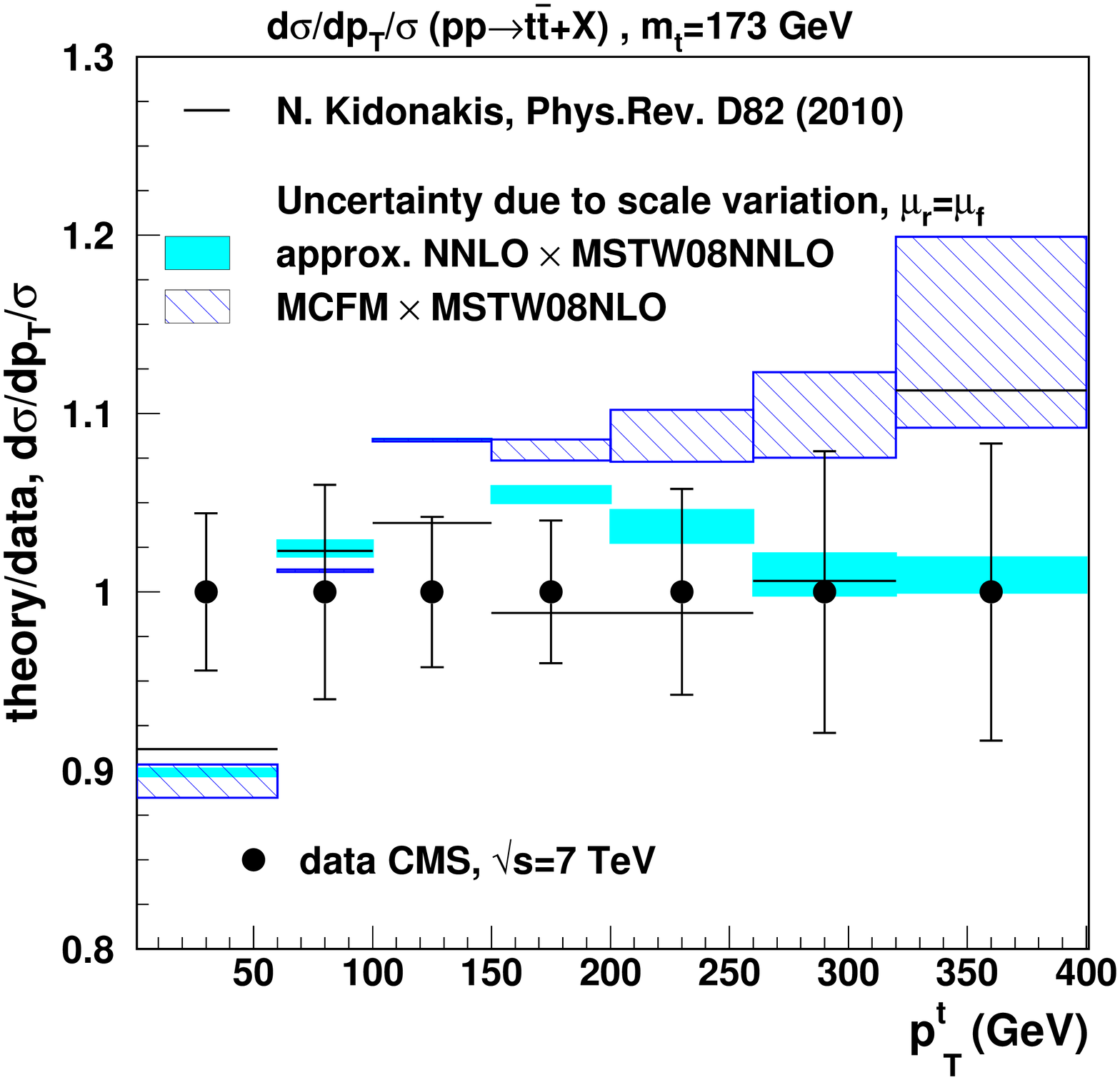}
\includegraphics[width=7cm, angle=0]{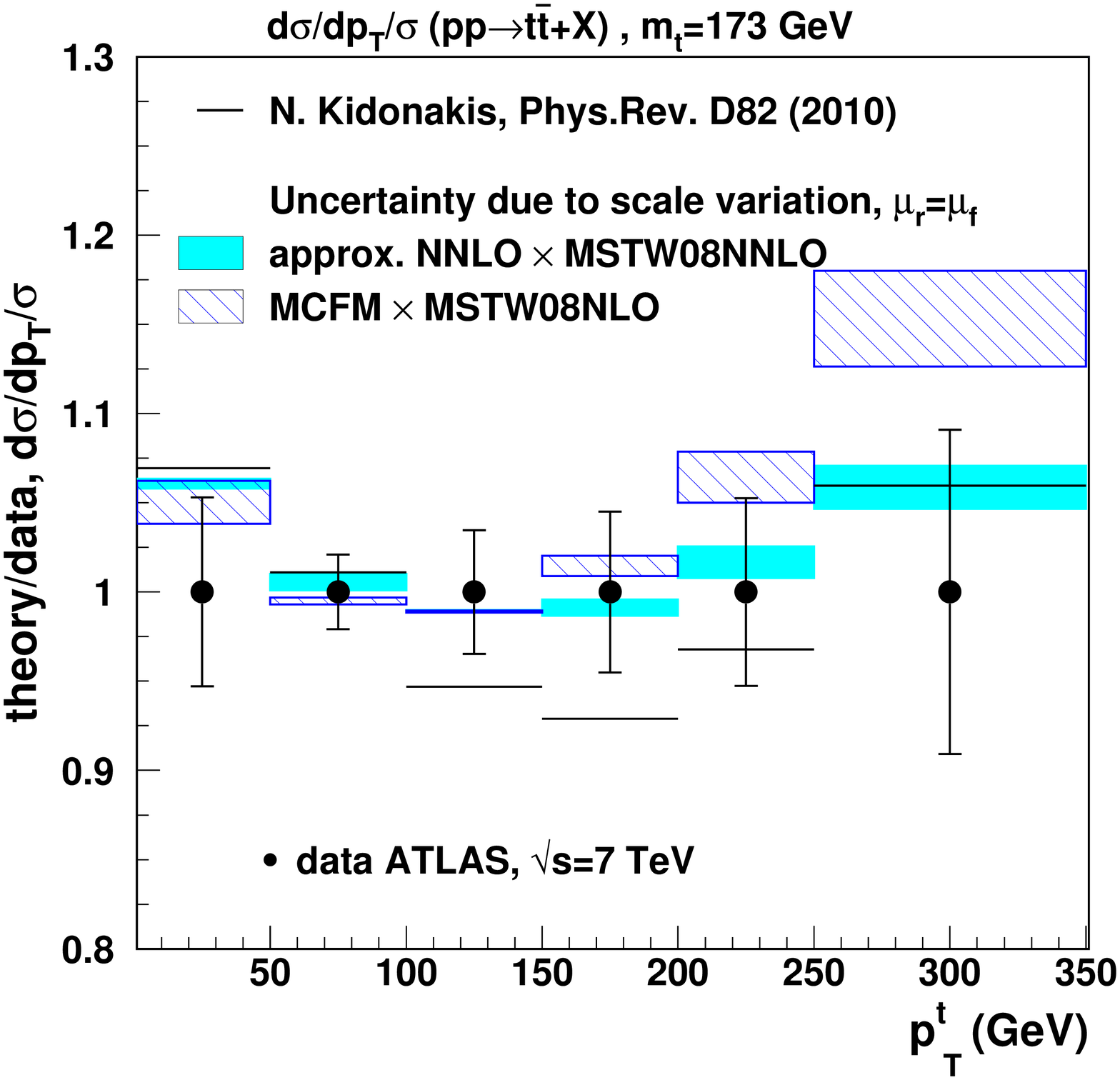}
\caption{Study of scale uncertainties for \textsc{MCFM} and \textsc{DiffTop} calculations 
for the top-quark $p^t_T$ distribution. 
Upper inset: absolute $p^t_T$ distribution at NLO with \textsc{MCFM} (hatched band), and approx. NNLO with \textsc{DiffTop} (shaded band). 
Lower inset: ratio of theory over data for CMS (left)~\cite{Chatrchyan:2012saa} and ATLAS (right)~\cite{Aad:2012hg} 
measurements. Here MSTW08 NLO (NNLO) PDFs are used for the \textsc{MCFM} (\textsc{DiffTop}) calculation.
Renormalization and factorization scales are set to $\mu_R=\mu_F=m_t$ and varied such as $m_t/2\leq \mu_R=\mu_F\leq 2m_t$. 
For comparison, the central prediction by N. Kidonakis (solid line) using MSTW08 NNLO PDFs is shown.
\label{mstw08-nlo-scale-unc}}
\end{center}
\end{figure}

The new LHC measurements give us the possibility of investigating the applicability of QCD 
factorization for hadronic cross sections with high precision.
In the past three decades QCD techniques have experienced 
stages of remarkable progress on theoretical and phenomenological ground.
In particular, theoretical tools were introduced to 
estimate the importance of the perturbative higher orders in cross-section calculations~\cite{Sterman:1986aj,Catani:1989ne,Catani:1990rp,Kidonakis:1997gm,Laenen:1998qw,Bonciani:1998vc}.
In Fig.~\ref{mstw08-nlo-scale-unc} we show the top-quark transverse momentum, $p^t_T$, where 
higher orders are important and have reduced scale dependence.

The non-unique separation between long- and short-distance dynamics 
of partons in the proton is understood in terms of factorization theorems,
in which inclusive hadronic cross sections are factorized 
in universal nonperturbative PDFs and fragmentation functions (FF), 
and hard scattering coefficients that can be computed in perturbative QCD.
These hard scattering functions
still contain, in certain kinematic regions, residuals of long-distance dynamic related to leftovers in the 
cancellation between real and virtual soft-gluon contributions.
These finite reminders limit the predictive power of QCD in the kinematic region 
close to the partonic production threshold where singular logarithmic 
terms give large corrections to the cross section.
Threshold resummation methods, which are based on the universality of 
the factorization in the soft and collinear limit, 
allow us to resum these large contributions to all orders 
so that the predictive power of pQCD is extended 
to the phase space regions of the partonic threshold.
By using threshold resummation methods one can derive approximate formulas 
at NNLO for differential distributions, in which the cross sections
are expanded in terms of the logarithmic enhanced contributions (appearing as plus distributions), 
and can therefore be written at various degrees of logarithmic accuracy.
Studies at the next-to-leading-logarithmic (NLL), next-to-next-to-leading-logarithmic (NNLL) 
accuracy and beyond can be found in Refs.~\cite{Kidonakis:2014isa,Kidonakis:2000ui,Kidonakis:2001nj,Kidonakis:2003tx,Kidonakis:2003qe,Kidonakis:2005kz,Kidonakis:2008mu,Kidonakis:2010tc,Kidonakis:2010dk,Czakon:2009zw} and references therein.

Recently, methods of heavy-quark effective theory (HQET) and soft-collinear effective 
theory (SCET) experienced enormous progress and received significant attention.
Such methods can be also used to write the factorization formula for the hard-scattering kernels in the threshold region
~\cite{Neubert:1993mb,Bauer:2000yr, Bauer:2001yt,Beneke:2002ph}.
Approximate NNLO calculations, based on SCET, for differential cross sections for heavy-quark production can be found in~\cite{Ahrens:2009uz,Ahrens:2010zv,Ahrens:2011mw,Ahrens:2011px,Ferroglia:2012uy,Ferroglia:2012ku,Ferroglia:2013awa} 
and references therein.
A study of differences and similarities between the traditional QCD threshold resummation formalism and 
SCET can be found in~\cite{Kidonakis:2011ca} for the $t\bar{t}$ case.

\section{Single particle kinematic \label{sec:Kinematics}}

Top-quark pairs are produced in hadronic reactions in which the scattering process is defined by 
\beq
H_1(P_1) + H_2(P_2)\rightarrow t(p_1) + \bar{t}(p_2) + X(k)\,,
\label{1PI_reac}
\eeq
where $H_1$ and $H_2$ are the incoming hadrons with momenta $P_1$ and $P_2$ respectively,
the final state top-quark has momentum $p_1$, the anti-top momentum $p_2$ and 
$X(k)$ represents any inclusive hadronic final state allowed by the reaction. 
At phenomenological level, the differential cross sections of interest are written in such a way that 
can be compared to measurements that are related to the detection of single particles or particle pairs.

When a single particle is detected in the final state, for instance the top-quark,
the one-particle inclusive (1PI) kinematic is used to determine the $p^t_T$ spectrum and 
the rapidity $y^t$ distribution of the top-quark. These distributions are obtained 
by integrating over the phase space of the anti-top 
(not observed) together with any real emission of radiation~\cite{Laenen:1998qw}.

When particle pairs are detected, the pair-invariant mass kinematic (PIM) 
is used to write a factorized cross section~\cite{Sterman:1986aj,Catani:1989ne,Kidonakis:1996aq,Kidonakis:1997gm,Bonciani:1998vc}
in terms of the invariant mass of the heavy system, collective rapidity, and additional variables like the angle between 
the final-state quark direction and the beam axis.
In what follows we will discuss differential distributions derived in the 
1PI kinematic case and will leave the PIM kinematic~\cite{Kidonakis:2001nj} 
for a forthcoming analysis\footnote{The extension of \textsc{DiffTop} 
to calculate approximate NNLO differential cross section for PIM kinematic is under development.}.

In the case of single particle detection, the reaction in Eq.(\ref{1PI_reac}) is written as 
\beq
H_1(P_1) + H_2(P_2)\rightarrow t(p_1) + X[\bar{t}](p'_2)\,,
\eeq
where $p'_2$ represents the recoil momentum.
The Mandelstam invariants at hadronic level are defined as 
\ba
S=(P_1+P_2)^2\,, ~~T_1=(P_2-p_1)^2 -m_t^2\,, ~~U_1=(P_1-p_1)^2-m_t^2\,, ~~S_4 = S + T_1 + U_1\,.
\ea
In the vicinity of the threshold the reaction is dominated by the following partonic subprocesses 
\ba
&&q(k_1) + \bar{q}(k_2) \rightarrow t(p_1) + X[\bar{t}](p'_2) \,,
\nonumber\\
&&g(k_1) + g(k_2) \rightarrow t(p_1) + X[\bar{t}](p'_2) \,,
\ea
where the initial-state parton momenta expressed in terms of momentum fractions $x_1, x_2$ 
are $k_1=x_1 P_1$ and $k_2=x_2 P_2$. The recoil momentum $p'_2=p_2 + k$   
accounts for momentum $p_2$ at the threshold and any additional radiation indicated by momentum $k$. 

In the vicinity of the partonic threshold,
the hadronic final state $X[\bar{t}(p'_2)]\equiv \bar{t}(p_2)$ and the anti-top carry momentum $p_2$. 
The Mandelstam invariants at parton level are defined as
\ba
&&s=x_1 x_2 S= (k_1+k_2)^2\,, ~t_1=x_2 T_1 =(k_2-p_1)^2 -m_t^2\,,  
\nonumber\\
&&u_1=x_1 U_1 =(k_1-p_1)^2-m_t^2\,, ~s_4 = s + t_1 + u_1\,,
\ea
where the inelasticity of the reaction is accounted for by the invariant $s_4={p'}_{2}^2-m_t^2$.

The factorized differential cross section is written as 
\ba
S^2\frac{d^2 \sigma(S,T_1,U_1)}{dT_1 ~dU_1} &=& \sum_{i,j=q,\bar{q},g} \int_{x_1^-}^{1} \frac{dx_1}{x_1}
\int_{x_2^-}^{1}\frac{dx_2}{x_2} f_{i/H_1}(x_1,\mu_F^2) f_{j/H_2}(x_2,\mu_F^2) 
\nonumber\\ 
&&\times \, \omega_{ij}(s,t_1,u_1,m_t^2, \mu_F^2,\alpha_s(\mu_R^2)) + {\cal O}(\Lambda^2/m_t^2)
\label{diffXsec}
\ea
where $f_{j/H}(x,\mu_F^2)$ is the probability of finding the parton $j$ in hadron $H$, 
$\mu_F$ and $\mu_R$ are the factorization and renormalization scales respectively, and  
$\omega_{ij}$ is the hard scattering cross section which depends on the kinematic of the reaction. 
Power suppressed terms $\Lambda^2/m_t^2$ are neglected here.
The integration limits in the factorization formula are given by 
\beq
x_1^-= -\frac{U_1}{S + T_1} \,, ~~ x_2^-= -\frac{x_1 T_1}{x_1 S + U_1}. 
\eeq
The double-differential cross section in Eq.(\ref{diffXsec}) can be expressed 
in terms of the transverse momentum $p^t_T$ of the top quark and its rapidity $y$ by observing that 
\beq
T_1= - \sqrt{s} ~m_T e^y\,,~~ U_1= - \sqrt{s} ~m_T e^{-y}\,,
\eeq
where the transverse mass $m_T$ is defined as $m_T=\sqrt{p_T^2 + m_t^2}$. 

According to QCD resummation, in the organization of the large logarithms 
at the threshold of the heavy system, the hard scattering $\omega_{ij}$ functions are expanded in 
terms of singular functions which are plus-distributions of the type
\footnote{This logarithmic structure refers to the 1PI kinematic. In PIM kinematic one has $\left[\ln^l{(1-z)}/(1-z)\right]_+$, where $z=M^2/s$ and $M$ is the invariant mass of the final-state heavy system.}
\beq
\left[\frac{\ln^l{(s_4/m_t^2)}}{s_4} \right]_{+} = \lim_{\Delta\rightarrow 0}\left\{\frac{\ln^l{(s_4/m_t^2)}}{s_4}\theta(s_4-\Delta)
+\frac{1}{l+1}\ln^{l+1}{\left( \frac{\Delta}{m_t^2}\right)}\delta(s_4)\right\}\,,
\eeq
where corrections are denoted as leading-logarithmic (LL) if $l=2 i+1$ at ${\cal O}(\alpha_s^{i+3})$ with $i=0,1,\dots$, 
as next-to-leading logarithm (NLL) if $l=2 i$, as next-to-next-to-leading logarithm (NNLL) if $l=2 i - 1$, and so on.

Since QCD threshold resummation calculations and resummed formulae 
for inclusive cross sections for heavy-quark pair production 
have been extensively discussed in the literature in the past years, 
we will omit explicit derivations 
unless necessary and limit ourselves to general 
definitions.
The calculation implemented in \textsc{DiffTop} strictly follows 
the derivation of Ref.~\cite{Kidonakis:2001nj} and references therein, 
which we refer to for details. Other particulars of this calculation can be found 
in~\cite{Kidonakis:2000ui,Kidonakis:2003tx,Kidonakis:2005kz,Kidonakis:2013zqa}.

\section{Overview of the threshold resummation calculation \label{sec:Overview}}

As described in \cite{Laenen:1998qw,Kidonakis:2001nj} the fully resummed expression for the hard scattering 
cross section of Eq.(\ref{diffXsec}) in 1PI kinematic is given by a trace in the color-tensor space 
of operators in the Mellin $N$-moment space 
\ba
&&\omega_{ij}[N,s,t_1,u_1,m_t^2,\mu_R^2,\mu_F^2,\alpha_s(\mu_R)] =
\textrm{Tr} \left\{ H_{ij}(s_4,t_1,u_1,m_t^2,\mu_R^2,\mu_F^2)
\right.
\nonumber\\
&&\left.
\times \overline{P}\exp \left[\int_{m_t}^{m_t/N} \frac{d\mu}{\mu} (\Gamma^{ij}_{S})^{\dagger}(\alpha_s(\mu))\right]
S_{ij}(s_4,t_1,u_1,m_t^2,\mu_R^2,\mu_F^2)
\right.
\nonumber\\
&&\left.
\times P\exp \left[\int_{m_t}^{m_t/N} \frac{d\mu}{\mu} \Gamma^{ij}_{S}(\alpha_s(\mu))\right]
\right\}
\exp \left[E_i(N_u,m_t,\mu_F,\mu_R)\right]\exp \left[E_j(N_t,m_t,\mu_F,\mu_R)\right]
\nonumber\\
&&\times \exp \left\{2\int_{\mu_R}^{m_t} \frac{d\mu}{\mu}\left[\gamma_i(\alpha_s(\mu))+\gamma_j(\alpha_s(\mu)) \right]\right\}\,,
\ea
where $N_u=N(-u_1/m_t^2)$, $N_t=N(-t_1/m_t^2)$ are the Mellin moments in 1PI kinematic. The functions
$H_{ij}=H_{ij}^{(0)}+(\alpha_s/\pi) H_{ij}^{(1)}+\cdots$ and $S_{ij}=S_{ij}^{(0)}+(\alpha_s/\pi) S_{ij}^{(1)}+\cdots$ 
are the hard and soft functions respectively. The functions $\Gamma_S=(\alpha_s/\pi) \Gamma_S^{(1)}+(\alpha_s/\pi)^2 \Gamma_S^{(2)}+\cdots$ 
are the soft anomalous dimension matrices which are path-ordered in $\mu$. 
Finally, $\gamma_i=(\alpha_s/\pi) \gamma_i^{(1)}+(\alpha_s/\pi)^2 \gamma_i^{(2)}+\cdots$ 
are the anomalous dimensions of the quantum field $i=q,g$. 
In our calculation, $\Gamma_S^{(2)}$ at two-loop for the massive case 
is given in~\cite{Becher:2009kw,Kidonakis:2009ev}. 

The exponentials $\exp{E_i}$ and $\exp{E_j}$ represent resummed expressions for the collinear and soft radiation 
from incoming and outgoing partons respectively, and are defined as
\ba
\exp \left[E_i(N_u,m_t,\mu_F,\mu_R)\right]&=&
\exp \left\{E_i(N_u, 2 k_i \cdot \zeta)\right\}
\nonumber\\
&\times& 
\exp \left\{-2\int_{\mu_R}^{2 k_i \cdot \zeta}\frac{d\mu}{\mu}\gamma_{i}(\alpha_s(\mu))
+2\int_{\mu_F}^{2 k_i \cdot \zeta}\frac{d\mu}{\mu}\gamma_{i/i}(N_u,\alpha_s(\mu))
\right\}\,,
\nonumber\\
\label{in_out_rad}
\ea
where $\gamma_{i/i}$ are the anomalous dimensions of the operator whose matrix element represents the parton density $f_{i/i}$ 
in the $\overline{\textrm{MS}}$ scheme. The vector $\zeta=p_2/m_t$ is used to define the distance from the threshold in 1PI kinematic:
$s_4/m_t^2\approx 2(\zeta \cdot k)/m_t$.
The first exponential in Eq.(\ref{in_out_rad}) is defined as 
\ba
E_i(N, 2 k_i \cdot \zeta)&=&\int_{0}^{\infty}dw \frac{(1-e^{-N w})}{w}
\nonumber\\
&\times&
\left\{\int_{w^2}^{1}\frac{d\lambda}{\lambda} A_i[\alpha_s(\sqrt{\lambda}~2 k_i \cdot \zeta)]+\frac{1}{2}\nu^i [\alpha_s(w~2 k_i \cdot \zeta)]
\right\}\,,
\ea
where functions $A_i$ and $\nu^i$ have perturbative expansion in $\alpha_s$ 
whose explicit expressions up to ${\cal O} (\alpha_s^2)$ can be found in~\cite{Kidonakis:2001nj,Kidonakis:2013zqa} 
for the Feynman and axial gauge.

After an $\alpha_s$-expansion of the resummed hard-scattering cross section, the general structure of the 
double-differential cross section at parton level is given by 
\ba
s^2 \frac{d\sigma_{ij}(s,t_1,u_1,\mu_R,\mu_F)}{dt_1~du_1} = \omega^{(0)}_{ij} (s_4,s,t_1,u_1)
+ \frac{\alpha_s}{\pi} \omega^{(1)}_{ij}(s_4,s,t_1,u_1) + \left(\frac{\alpha_s}{\pi}\right)^2  \omega^{(2)}_{ij}(s_4,s,t_1,u_1)\,,
\nonumber\\
\ea
where $\omega^{(0)}_{ij}$ is the cross section at the Born level and the one-loop soft-gluon correction beyond the LL approximation 
can be written as
\ba
\omega^{(1)}_{ij}(s_4,s,t_1,u_1) &=& C_1^{(1)} \left[\frac{\ln(s_4/m_t^2)}{s_4}\right]_+
+ \left(C_0^{(1)}   + C_{0,\mu_F}^{(1)}\right)\ln\left(\frac{\mu_F^2}{m_t^2}\right) \left[\frac{1}{s_4}\right]_+ 
\nonumber\\
&+& \left[R_1 +r_{1,\mu_F}\ln\left(\frac{\mu_F^2}{m_t^2}\right)
+ r_{1,\mu_R}\ln\left(\frac{\mu_R^2}{m_t^2}\right)\right] ~\delta(s_4)\,,
\ea
where coefficients $C_0^{(1)},C_1^{(1)}$ (in which we suppress the $ij$ indices) 
and $r_1$ can be found in Appendix B of~\cite{Kidonakis:2001nj}.
At the NNLL accuracy the coefficient $R_1$ which consists of virtual graph and soft-gluon radiation contributions, 
contains the NLO matching term $\textrm{Tr}\left[H^{(0)}S^{(1)}+H^{(1)}S^{(0)}\right]$.
The expression of the NLO matching term is that of the soft plus virtual ($S+V$) contributions in Eq.(6.19) 
of Ref.~\cite{Beenakker:1988bq} for the $gg$ channel, 
and that of the contributions in Eq.(4.7) of Ref.~\cite{Beenakker:1990maa} for the $q\bar{q}$ channel. 
The Coulomb interactions, due to gluon exchange between 
the final-state heavy quarks, are included at 1-loop level.

The two-loop corrections at the NNLL accuracy can be written as 
\ba
&&\omega^{(2)}_{ij}(s_4,s,t_1,u_1) = C_3^{(2)} \left[\frac{\ln^3(s_4/m_t^2)}{s_4}\right]_+ 
+ \left[C_2^{(2)} + C_{2,\mu_F}^{(2)}\ln\left(\frac{\mu_F^2}{m_t^2}\right)\right]\left[\frac{\ln^2(s_4/m_t^2)}{s_4}\right]_+ 
\nonumber\\
&&+ \left[C_1^{(2)} + C_{1,\mu_F}^{(2)}\ln\left(\frac{\mu_F^2}{m_t^2}\right) 
+ C_{1,\mu_R}^{(2)}\ln\left(\frac{\mu_R^2}{m_t^2}\right) + \overline{C}_{1,\mu_F}^{(2)}\ln^2\left(\frac{\mu_F^2}{m_t^2}\right)\right]
\left[\frac{\ln(s_4/m_t^2)}{s_4}\right]_+ 
\nonumber\\
&&+ \left[
C_0^{(2)} + C_{0,\mu_F}^{(2)}\ln\left(\frac{\mu_F^2}{m_t^2}\right) 
+ C_{0,\mu_R}^{(2)}\ln\left(\frac{\mu_R^2}{m_t^2}\right) 
\right.
\nonumber\\
&&\left.
~~~~+ \overline{C}_{0,\mu_F}^{(2)}\ln^2\left(\frac{\mu_F^2}{m_t^2}\right)
+ \overline{C}_{0,\mu_F,\mu_R}^{(2)}\ln\left(\frac{\mu_F^2}{m_t^2}\right)\ln\left(\frac{\mu_R^2}{m_t^2}\right)
\right]
\left[\frac{1}{s_4}\right]_+ 
\nonumber\\
&&+\left[R_2  + r_{2,\mu_R}\ln\left(\frac{\mu_R^2}{m_t^2}\right) 
+ r_{2,\mu_F}\ln\left(\frac{\mu_F^2}{m_t^2}\right) 
+ r_{2,\mu_F,\mu_R}\ln\left(\frac{\mu_F^2}{m_t^2}\right)\ln\left(\frac{\mu_R^2}{m_t^2}\right)
\right.
\nonumber\\
&&\left.
~~~~+ \overline{r}_{2,\mu_R}\ln^2\left(\frac{\mu_R^2}{m_t^2}\right) 
+ \overline{r}_{2,\mu_F}\ln^2\left(\frac{\mu_F^2}{m_t^2}\right) 
\right]~\delta(s_4)\,,
\ea
where again, explicit expressions for the coefficients $C_{0,\mu_F}^{(2)},C_{1}^{(2)},\dots$ can be found in Appendix B of~\cite{Kidonakis:2001nj}.
In the current implementation the scale-independent coefficient $C_0^{(2)}$ contains the contribution from 
the soft anomalous dimension $\Gamma_{S}^{(2)}$ given in~\cite{Becher:2009kw,Kidonakis:2009ev} 
plus process-independent terms~\cite{Kidonakis:2003qe,Kidonakis:2013zqa}
which are universal in the $q\bar{q}$ and $gg$ channels respectively.
These contributions are formally at the next-to-next-to-next-leading logarithmic (NNNLL) accuracy.
At this level of accuracy, $C_0^{(2)}$ is therefore not exact due to incomplete separation of the individual color structures.
From a comparison with the analogous approximation for the inclusive cross
section we expect only small numerical deviations for the exact result.
In the scale-independent coefficient $R_2$ we only include NNNLL subleading terms 
coming from moment to momentum space inversion~\cite{Kidonakis:2000ui}.
The full knowledge of $R_2$ requires matching conditions at NNLO which include explicit analytical 
expressions for $H^{(2)}$ and $S^{(2)}$.
In the given kinematics (1PI), these are currently not available for $H^{(2)}$ and $S^{(2)}$ 
and those terms are thus set to zero.
The current uncertainty on $R_2$ is therefore the dominant source.

\begin{figure}[ht]
\begin{center}
\includegraphics[width=5.5cm, angle=-90]{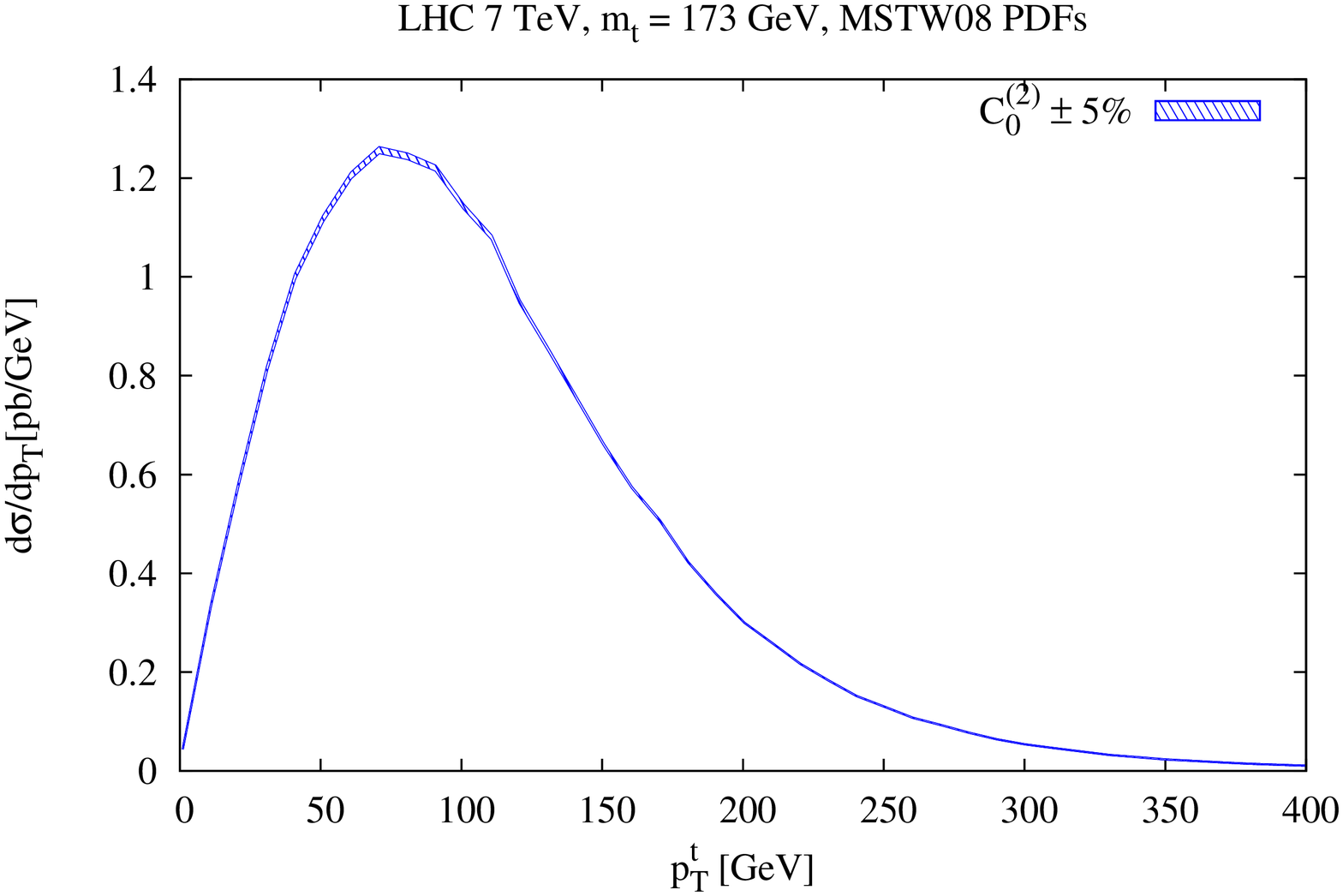}
\includegraphics[width=5.5cm, angle=-90]{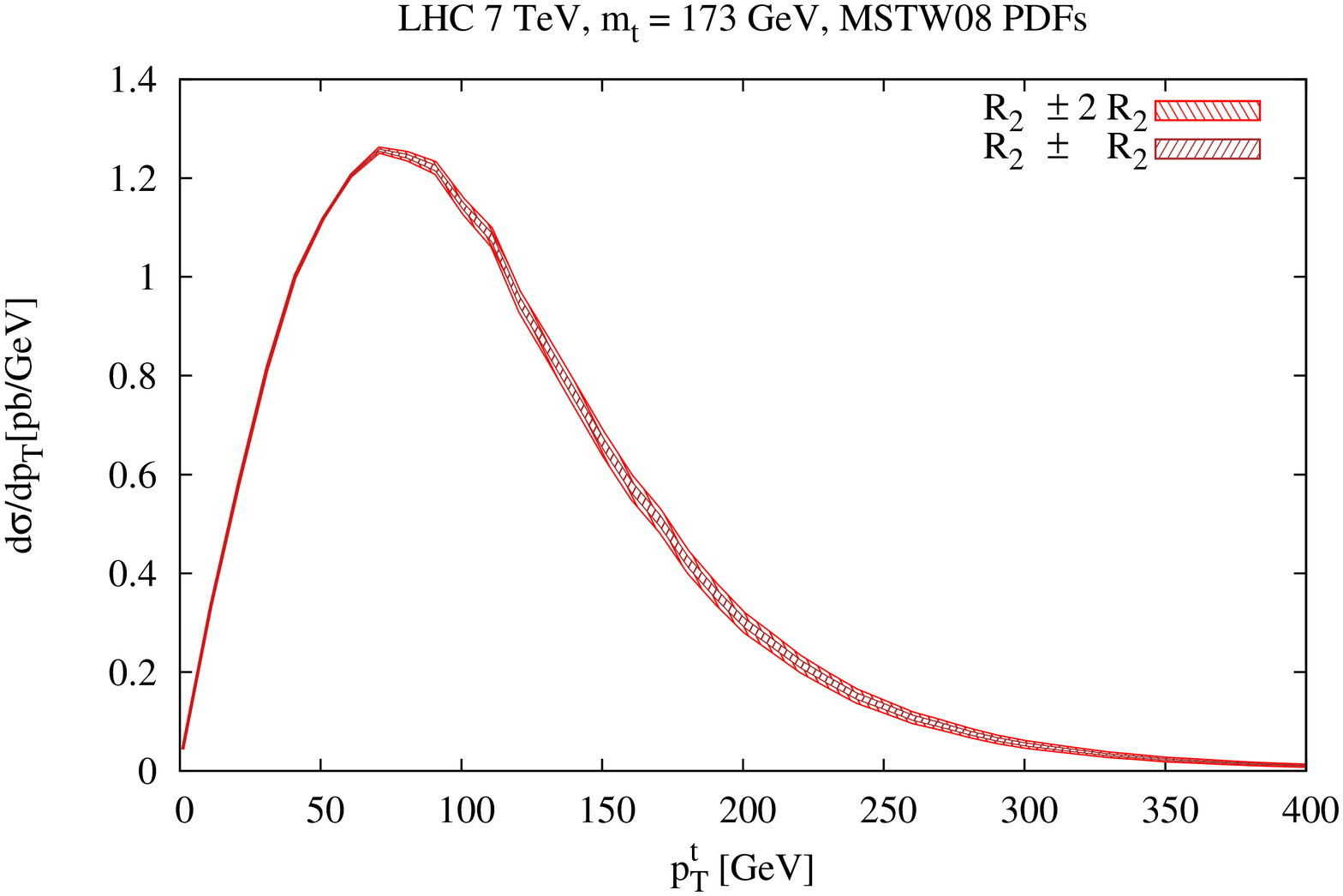}
\caption{Left: Top-quark $p_T$ spectrum in which the coefficient $C^{(2)}_0$ is varied within its 5\% 
while $R_2$ is kept fixed. Right: here the coefficient $R_2$ is 
varied by adding and subtracting $2 R_2$ while $C^{(2)}_0$ is kept fixed.
\label{C2-R2var}}
\end{center}
\end{figure}

To give an estimate of the dependence of the $p_T$ spectrum on the coefficients $C^{(2)}_0$ and $R_2$ 
that are known only partially in our computation, we generated a set of predictions in which these coefficients are 
varied. In Fig.~\ref{C2-R2var} we illustrate these variations separately, where the coefficient 
$C^{(2)}_0$ is varied within 5\% of its magnitude by keeping $R_2$ as fixed, and $R_2$ is varied 
by adding and subtracting $R_2$ or $2 R_2$, and $C^{(2)}_0$ is fixed. 
Variations of $C^{(2)}_0$ produce modifications of magnitude and shape of the $p_T$ spectrum which are more 
pronounced in the peak region, where these are approximately 1\% when $R_2$ is fixed. 
Being $R_2$ the less known contribution, we allow this coefficient to vary in 
a much larger interval to be more conservative.  
Variations of $R_2$ within its 5\% when $C^{(2)}_0$ is fixed, are found to be negligible.
In Fig.~\ref{C2-R2var} larger variations of $R_2$ have an impact over all the $p_T$ range beyond the peak, 
where modifications of the magnitude and shape can be approximately 7-10\% or more at $p_T\approx$ 200 GeV and are 
larger than those obtained by a simultaneous variation of factorization and renormalization scales.
The uncertainties relative to $C^{(2)}_0$ and $R_2$ are part of the systematic uncertainty associated
to approximate calculations of this kind which are based on threshold expansions.  
Explicit expressions in SCET for $S^{(2)}$ in PIM kinematic, in the limit of boosted top quarks, can be found in Ref.~\cite{Ferroglia:2012uy}. 
Related studies are presented in Refs.~\cite{Mitov:2009sv,Mitov:2010xw,Ferroglia:2009ep,Ferroglia:2009ii,Ferroglia:2012uy,Czakon:2013hxa}.

By setting $\mu_R = \mu_F = \mu$ one can write the inclusive total 
partonic cross section in terms of scaling functions
$f_{ij}^{(k,l)}$ that are dimensionless and depend only on the variable $\eta=s/(4 m_t^2) - 1$
\ba
\sigma_{ij}(s,m_t^2,\mu^2) = \frac{\alpha_s^2(\mu)}{m_t^2} \sum_{k=0}^{\infty} \left(4\pi \alpha_s(\mu)\right)^k
~\sum_{l=0}^{k}f_{ij}^{(k,l)}(\eta)\ln^l\left(\frac{\mu^2}{m_t^2}\right).
\ea
To reduce the impact of threshold logarithms 
in the large-$\eta$ regions of the scaling functions we 
made use of dumping factors as it is done in Ref.~\cite{Meng:1989rp}.

\section{Sensitivity of $t\bar{t}$ production to PDF-related aspects of QCD. \label{sec:pheno}}

This section addresses details of the phenomenological analysis of the uncertainties on the predictions
for differential $t\bar{t}$ production cross sections at the LHC. In particular, PDF uncertainties 
are studied and compared to the current experimental precision of the measurements.
The approximate NNLO \textsc{DiffTop} predictions, obtained by using different PDF sets, 
are confronted to the recent measurements of differential distributions for $t\bar{t}$ production 
at $\sqrt{S}=7$ TeV by the CMS~\cite{Chatrchyan:2012saa} and ATLAS~\cite{Aad:2014zka,TheATLAScollaboration:2013eja} 
collaborations. The theoretical systematic uncertainties associated to variations of PDFs, $\alpha_s(M_Z)$, 
scale, and $m_t$, are investigated individually. In particular, transverse momentum $p^{t}_{T}$ and 
rapidity $y^t$ distributions of the final state top-quark, measured by CMS, and the $p^{t}_{T}$ distribution 
by  ATLAS\footnote{In the data set we considered, $y^t$ is not provided by the ATLAS collaboration.} are studied.
The on-shell pole-mass definition for the top-quark mass is used, and the value 
of $m_t=173$ GeV is chosen. The scales are set to $\mu_F=\mu_R=m_t$.
The experimental measurements published by CMS and ATLAS are differential 
distributions that are normalized to the total cross section in bins of $p^t_T$ and $y^t$. This representation 
of experimental data is motivated by (partial) cancellation of systematic uncertainties. 

\begin{figure}[ht]
\begin{center}
\includegraphics[width=7cm, angle=0]{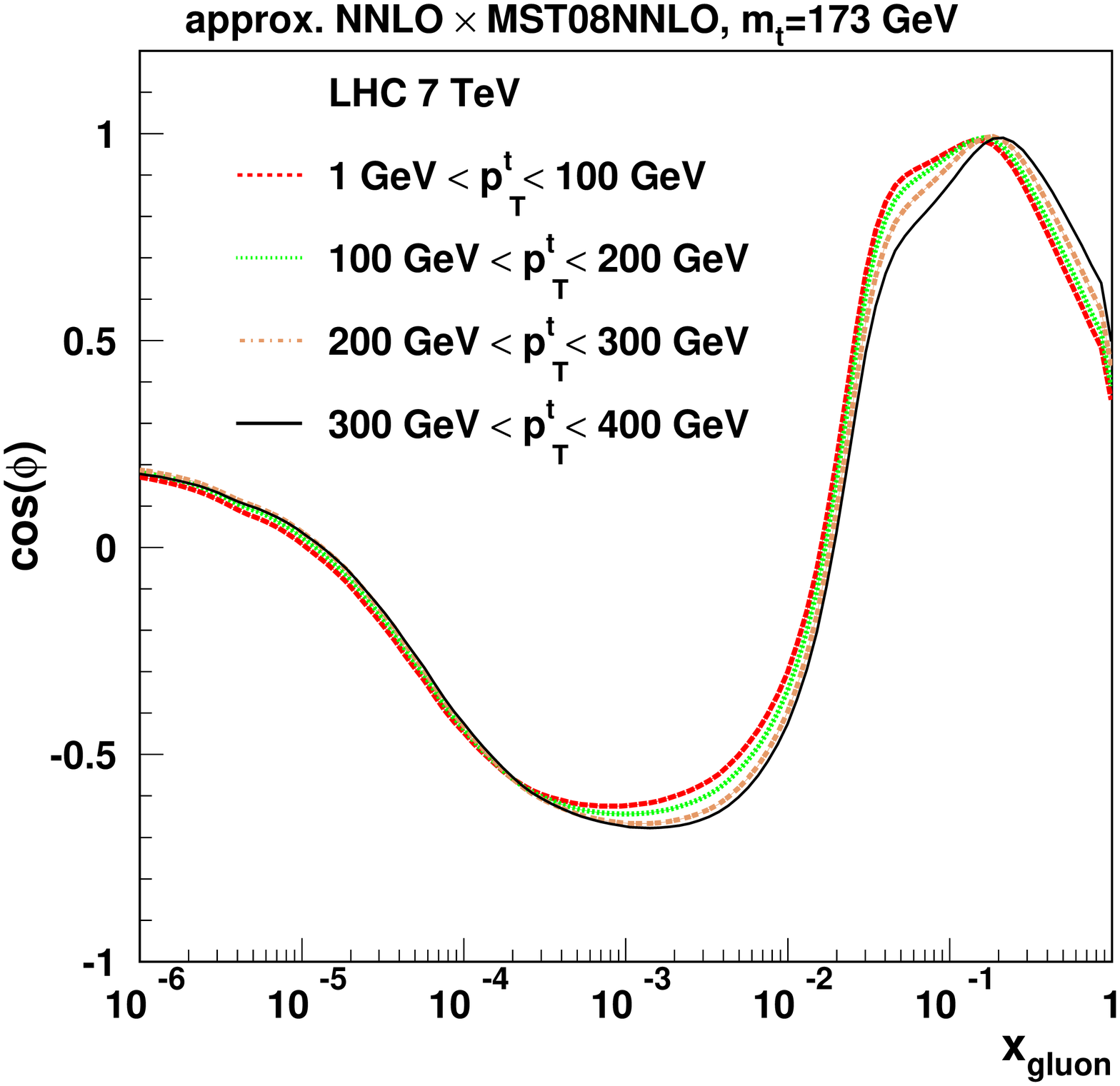}
\includegraphics[width=7cm, angle=0]{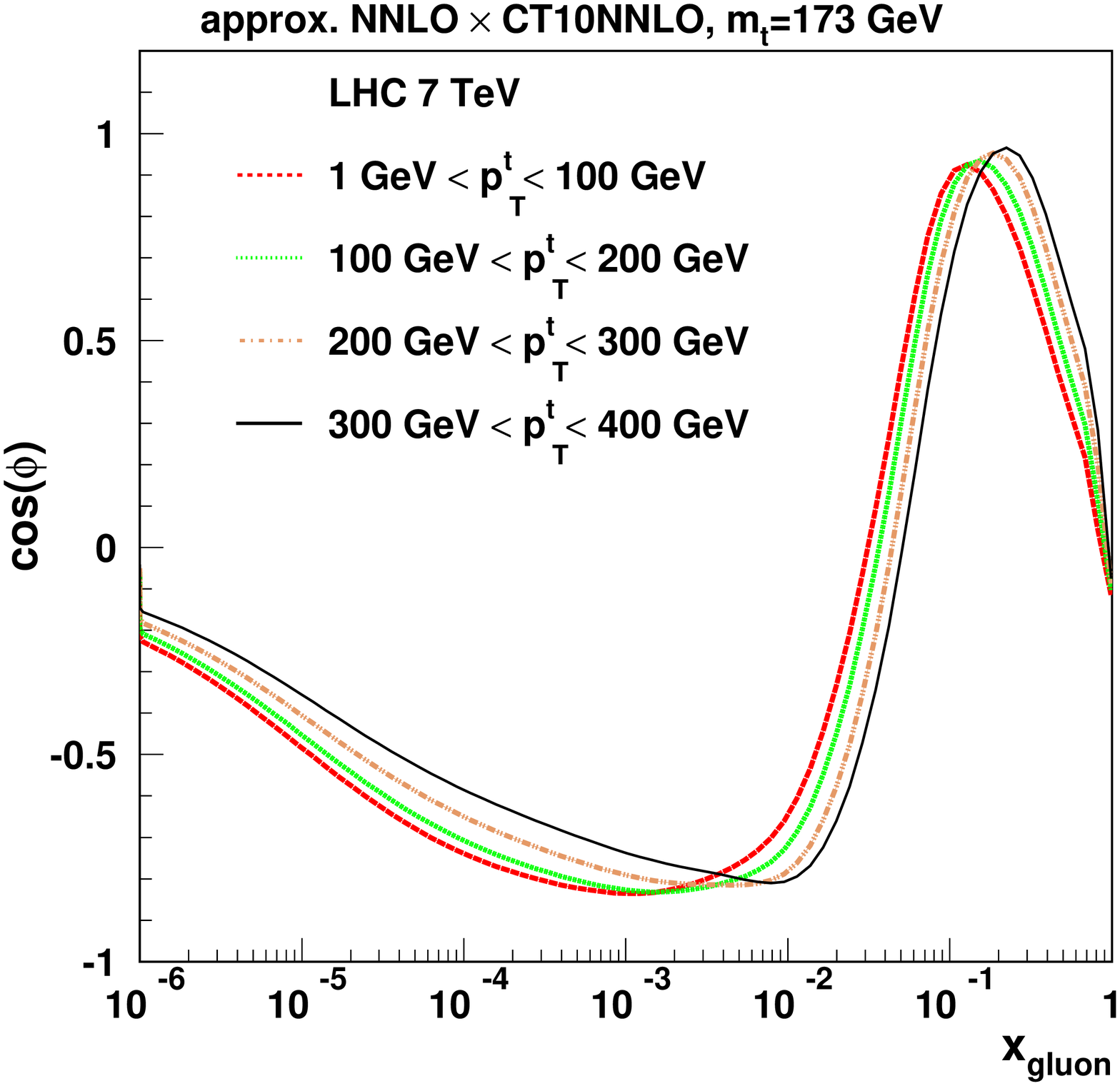}
\caption{Theoretical correlation cosine as a function of $x_\textrm{gluon}$
for the $p^{t}_{T}$ distribution in $t\bar{t}$ production at the LHC at $\sqrt{S}=7$ TeV. 
MSTW08 (left) and CT10 (right) PDFs at NNLO are used. Different lines represent each $p^{t}_{T}$ bin.
\label{corr-cos}}
\end{center}
\end{figure}

The correlation between the $p^t_T$ distribution for $t\bar{t}$ production at the LHC at $\sqrt{S}=7$ TeV 
and the gluon, as a function of $x$ of the gluon ($x_{\textrm{gluon}}$) is illustrated in Fig.~\ref{corr-cos}. 
A strong correlation is observed at $x_{\textrm{gluon}} \geq 0.01$. Here, the $p^{t}_{T}$ distribution is 
averaged in 4 bins and the correlation cosine $\cos{\phi}$, as defined in~\cite{Nadolsky:2008zw}, is 
evaluated for each bin. The predictions using MSTW08~\cite{Martin:2009iq} (left) and CT10~\cite{Gao:2013xoa} (right) 
PDFs at NNLO are shown here, while those using ABM11~\cite{Alekhin:2012ig}, HERAPDF1.5~\cite{CooperSarkar:2011aa}, 
and NNPDF2.3~\cite{Ball:2012cx} have similar behaviors.

The uncertainties associated to the various PDFs are computed by using the prescription given by each PDF group.
All envelopes represent the 68\% confidence level (CL). For the CT10 case, being the CT10NNLO PDF given at 90\% CL, 
the asymmetric PDF errors are rescaled by a factor 1.642.
In the ABM11 case, the total uncertainty obtained by using the symmetric formula for the eigenvector sets, 
represents the PDF + $\alpha_s$ uncertainty at the 68\% CL.
The uncertainty of HERAPDF1.5 NNLO is determined by including the experimental (at the 68\% CL), 
model and parametrization uncertainties, which are summed in quadrature to obtain the total error.

The uncertainty associated to the variations of $\alpha_s(M_Z)$ is computed by using 
the central $\alpha_s(M_Z)$ values given by each PDF group, 
and by considering up- and down-variations $\Delta\alpha_s(M_Z)=\pm 0.001$.
This is more conservative with respect to the 68\% CL variation $\Delta\alpha_s(M_Z)=0.0007$, 
reported in the PDG 2012~\cite{Beringer:1900zz}. 
The central values for $\alpha_s(M_Z)$ provided by the different PDF groups 
are: 0.1134, 0.118, 0.1176, 0.1171, 0.118 for ABM11~\cite{Alekhin:2012ig}, CT10~\cite{Gao:2013xoa}, HERAPDF1.5~\cite{CooperSarkar:2011aa}, 
MSTW08~\cite{Martin:2009iq}, and NNPDF2.3~\cite{Ball:2012cx}, respectively.
The HERAPDF1.5 $\alpha_s(M_Z)$ uncertainty is obtained by varying $0.1170\leq \alpha_s(M_Z) \leq 0.1190$, 
resulting in a larger uncertainty on the cross section, as compared to the
other PDF sets\footnote{The variation corresponding to the inclusion 
of members 9-11 of the HERAPDF1.5 $\alpha_s$ sets gives an increased uncertainty of approximately $\pm 6.3\%$.}. 

The uncertainty related to the choice of the scale has been estimated by varying $m_t/2 \leq \mu_R=\mu_F\leq 2m_t$. 
As shown in Fig.~\ref{mstw08-nlo-scale-unc}, where the approximate NNLO prediction 
and the full NLO calculation obtained by \textsc{MCFM}~\cite{Campbell:2000bg} are compared,
the reduction of the scale dependence is substantial.
When these theory predictions are compared to the LHC data for the $p^t_T$ distribution, 
the shape is also modified when passing from NLO to the approximate NNLO. 
In the approximate NNLO case, the theoretical description of the measurements is significantly improved. 

The uncertainty associated to the pole mass $m_t$ has been assessed by considering variations 
$\Delta m_t=\pm 1$ GeV around the central value $m_t=173$ GeV. 

\subsection{Cross sections of $t\bar{t}$ production at approximate NNLO obtained with different PDFs.\label{sec:numeric}}
In Tab.~\ref{tab1-LHC7} we summarize the results for the $t\bar{t}$ total inclusive cross section at the LHC 
at $\sqrt{S}=$7 TeV for each PDF set with relative uncertainties.
\begin{table}[h]
%\hspace{-6.0cm}
\begin{centering}
\begin{scriptsize}
\begin{tabular}{|c|c|c|c|c|c|c|}
\hline 
\multicolumn{6}{|c|}{LHC 7 TeV~ $m_t=173$ GeV}\tabularnewline
\hline 
PDF set & $\sigma_{t\bar{t}}$ [pb] & $\delta_{PDF}$ [pb]  & $\delta_{\alpha_s}$ [pb] & $\delta_{scale}$ [pb] & $\delta_{m_t}$ [pb] \\
\hline 
ABM11 & $140.9^{+7.6}_{-8.2}$ & $^{+4.5\%}_{-4.5\%}$ &  $-$ & $^{+0.0\%}_{-1.3\%}$  & $^{+2.8\%}_{-3.4\%}$\\
\hline
CT10 & $180.3^{+11.3}_{-10.6}$ & $^{+4.8\%}_{-3.9\%}$ &  $^{+2.5\%}_{-2.5\%}$ & $^{+1.8\%}_{-2.4\%}$ & $^{+2.8\%}_{-3.3\%}$\\
\hline 
HERA1.5 & $185.2^{+40.4}_{-14.3}$ & $^{+2.3\%}_{-3.6\%}(exp)~ ^{+3.7\%}_{-0.0\%}(param.) ~^{+21.2\%}_{-3.1\%} (mod.)$ &  $^{+3.2\%}_{-3.2\%}$ & $^{+1.7\%}_{-1.3\%}$  & $^{+2.7\%}_{-3.2\%}$\\
\hline 
MSTW08 & $179.4^{+8.9}_{-9.8}$ & $^{+2.7\%}_{-2.8\%}$ &  $^{+2.4\%}_{-2.4\%}$ & $^{+1.8\%}_{-2.4\%}$  & $^{+2.8\%}_{-3.2\%}$\\
\hline 
NNPDF23 & $179.9^{+8.9}_{-10.1}$ & $^{+3.0\%}_{-3.0\%}$ & $^{+2.0\%}_{-2.0\%}$ & $^{+1.8\%}_{-2.6\%}$  & $^{+2.8\%}_{-3.3\%}$\\
\hline 
\end{tabular}
\end{scriptsize}
\par
\end{centering}
\caption{Values of the total inclusive $t\bar{t}$ cross section with corresponding uncertainties for PDFs, 
$\alpha_s(M_Z)$, scale dependence and $m_t$. \label{tab1-LHC7}}
\end{table}

In Tab.~\ref{tab2-LHC7} we report for comparison the results obtained by using the \textsc{Top++}~\cite{Czakon:2011xx} code 
with the same input configuration, where the scale uncertainty is obtained by varying $\mu_F$ and $\mu_R$ independently.
\textsc{DiffTop} predictions are larger than those of \textsc{Top++} by approximately 6-6.5\% in the fixed-order case, and 3-3.5\% in the resummed case. 
In the fixed-order full NNLO calculation for the inclusive cross section, a simultaneous variation of $\mu_F$ and $\mu_R$ gives already 
the full scale variation and all independent variations are included in the envelope $1/2 \leq  \mu_F=\mu_R \leq 2$.
\textsc{DiffTop} scale uncertainty for simultaneous scale variation is underestimated 
because of missing contributions in the coefficients 
$R_2$ and $C_0^{(2)}$, in particular $R_2$, which has impact on the large $p_T$-spectrum.  
Also, missing contributions from the $qg$ channel play a role at higher orders at large $p_T$.
These missing contributions spoil the agreement in the case of inclusive observables such as the total cross section. 
However, the local description of differential observables can still be approximated sufficiently 
well by threshold expansions in regions that are not strongly affected by hard gluon radiation. 
For example in the region $1 \lesssim p_T \lesssim  250-300$ GeV, where the bulk of the data is currently given.    
Exact predictions for the fixed-order NLO differential and total inclusive cross sections are 
in good agreement with the approximate NLO ones (see for example Ref.~\cite{Guzzi:2013noa}) where the known one-loop 
soft and hard functions are included. This is an indication of the fact that the approximate NNLO predictions can be improved  
once the two-loop soft and hard functions from the full NNLO will be available and included in the calculation.
A crude estimate of the systematic uncertainties associated to the missing contributions 
in functions $R_2$ and $C_0^{(2)}$ for the total and differential cross sections was given 
in Sec.~\ref{sec:Overview}. A more realistic estimate of such uncertainties (variations can actually be combined)
requires a separate analysis on which work is in progress and it will be addressed in a forthcoming paper by the authors. 
In the following figures these uncertainties are therefore not included. 
\begin{table}[h]
%\hspace{-6.0cm}
\begin{centering}
\begin{scriptsize}
\begin{tabular}{|c|c|c|c|c|}
\hline 
\multicolumn{5}{|c|}{\textsc{Top++} scale dependence at the LHC 7 TeV~ $m_t=173$ GeV}\tabularnewline
\hline 
PDF set & $\sigma_{t\bar{t}}^{NNLO}$ [pb] & LL & NLL & NNLL \\
\hline 
ABM11    & $134.0^{+5.1(3.8\%)}_{-8.3(6.2\%)}$  & $137.1^{+4.6(3.4\%)}_{-8.2(6.0\%)}$ & $138.2^{+3.2(2.4\%)}_{-4.9(3.6\%)}$ & $138.1^{+3.5(2.5\%)}_{-4.4(3.2\%)}$ \\
\hline
CT10     & $169.1^{+6.8(4.0\%)}_{-10.9(6.4\%)}$ & $173.0^{+6.1(3.5\%)}_{-10.6(6.1\%)}$ & $174.3^{+4.3(2.5\%)}_{-6.7(3.9\%)}$ & $174.2^{+4.6(2.7\%)}_{-6.0(3.4\%)}$ \\
\hline 
HERA1.5  & $173.6^{+6.6(3.8\%)}_{-9.4(5.4\%)}$ & $177.6^{+5.9(3.3\%)}_{-9.1(5.1\%)}$ & $178.9^{+2.2(1.3\%)}_{-3.1(1.7\%)}$& $178.8^{+4.8(2.7\%)}_{-4.2(2.3\%)}$\\
\hline 
MSTW08   & $168.6^{+6.7(4.0\%)}_{-10.8(6.4\%)}$  & $172.5^{+6.0(3.5\%)}_{-10.5(6.1\%)}$ & $173.7^{+4.2(2.4\%)}_{-6.6(3.8\%)}$ & $173.6^{+4.5(2.6\%)}_{-5.9(3.4\%)}$  \\
\hline  
NNPDF23  & $169.1^{+7.0(4.2\%)}_{-11.1(6.6\%)}$  & $173.1^{+6.3(3.7\%)}_{-11.0(6.3\%)}$  & $174.4^{+4.3(2.4\%)}_{-6.8(3.9\%)}$  & $174.3^{+4.6(2.6\%)}_{-6.0(3.5\%)}$  \\
\hline 
\end{tabular}
\end{scriptsize}
\par
\end{centering}
\caption{\textsc{Top++} (ver 2.0) values for the total inclusive $t\bar{t}$ 
cross section at full NNLO without and with the inclusion of threshold resummation, 
with scale dependence uncertainty obtained 
by using independent variations of $\mu_F$ and $\mu_R$. \label{tab2-LHC7}}
\end{table}

\begin{figure}[h]
\begin{center}
\includegraphics[width=7.cm, angle=0]{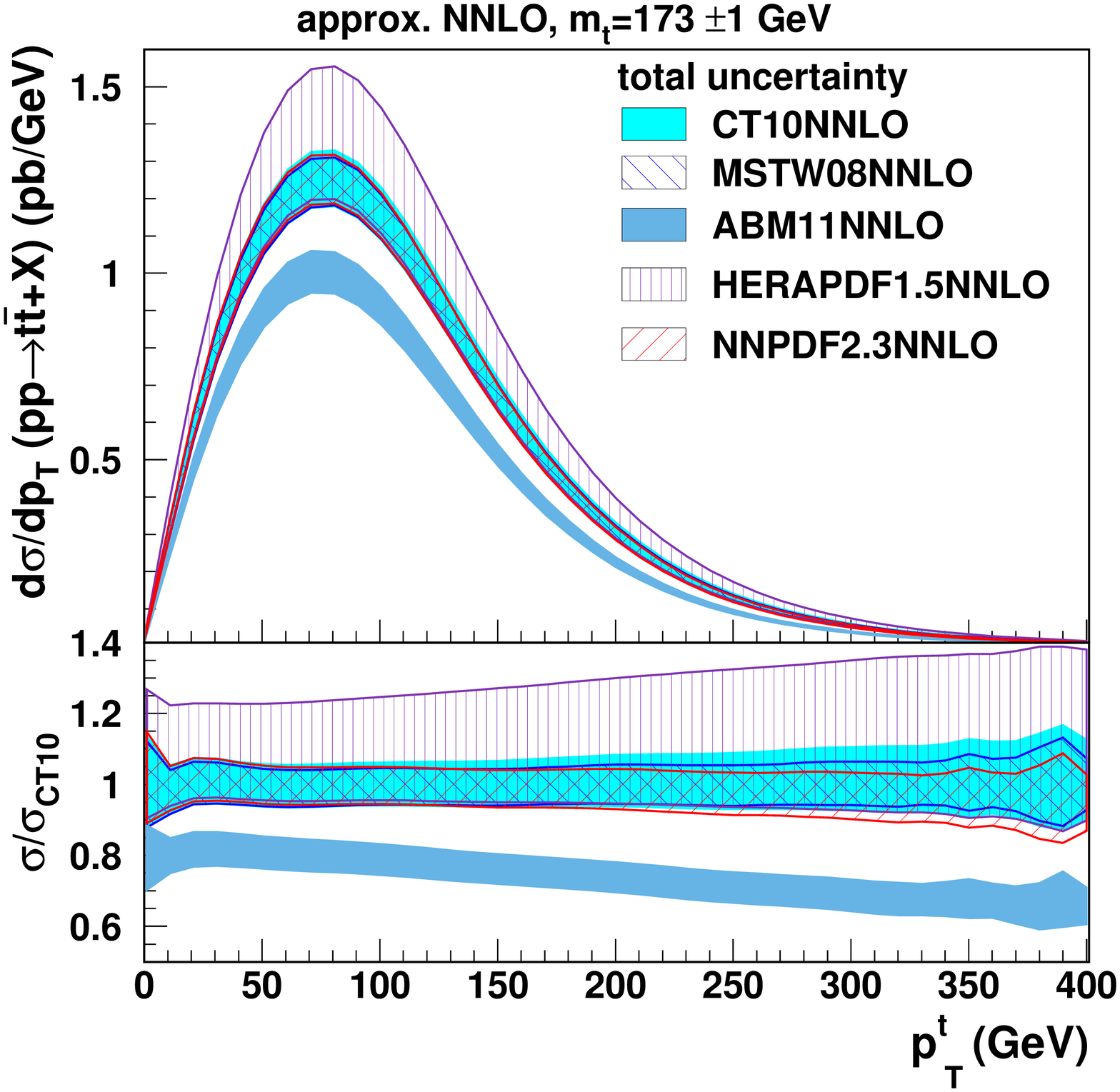}
\includegraphics[width=7.cm, angle=0]{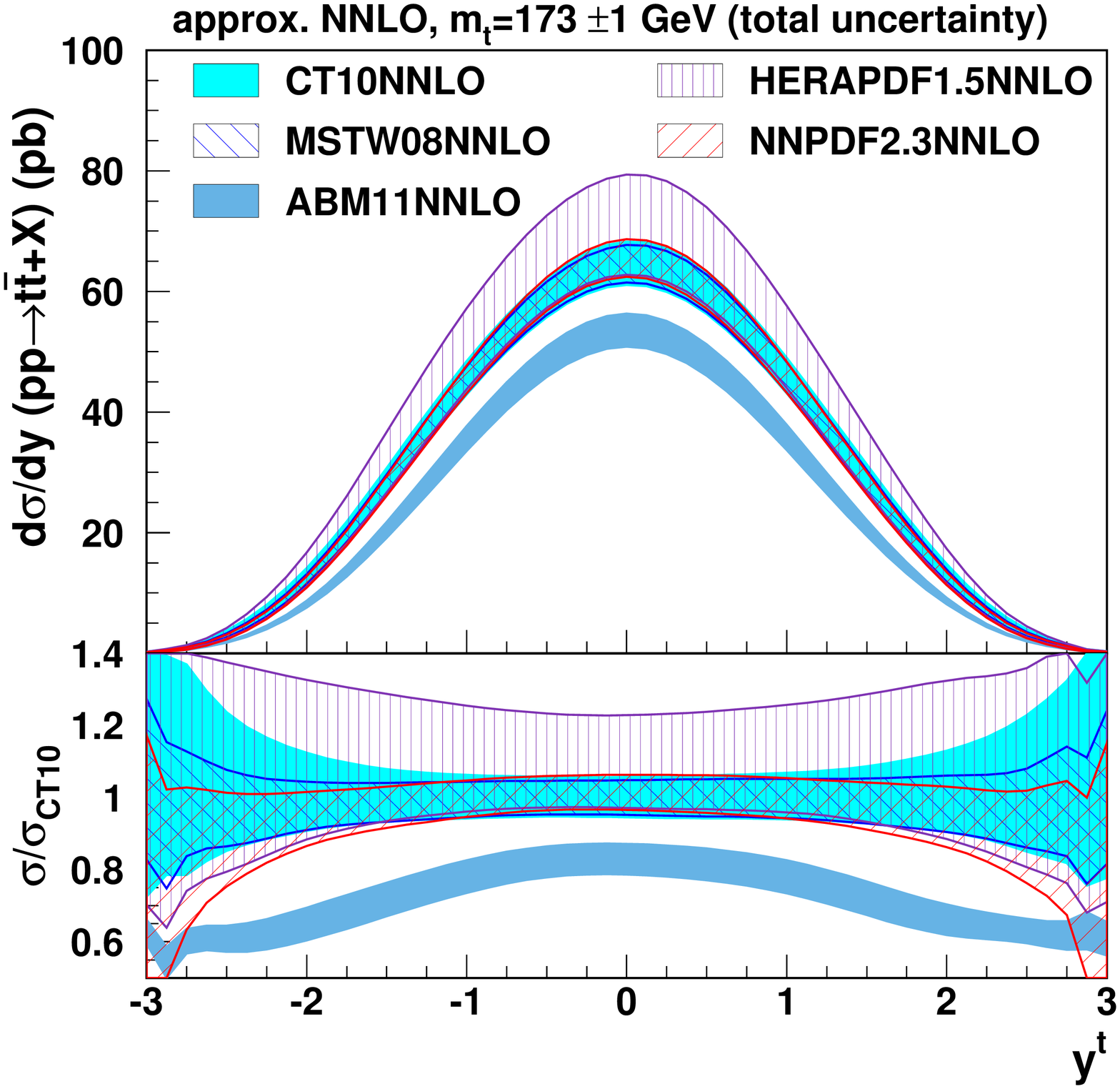}
\end{center}
\caption{The approximate NNLO predictions for top-quark pair production cross sections at the LHC at $\sqrt{S} = 7$ TeV as functions of $p^t_T$(left) and $y^t$(right). 
Predictions obtained by using different PDF sets are 
presented by bands of different hatches. The total uncertainty is obtained by summing the uncertainties due to 
PDFs, $\alpha_s$, $m_t$ and scale variations in quadrature. 
\label{allPDFs-unc-pt}}
\end{figure}

In Fig.~\ref{allPDFs-unc-pt} predictions for the absolute $p^t_T$ and $y^t$ distributions 
are shown, with all used PDF sets that are compared within the respective total uncertainties. 
The spread of the central values is mostly due to the fact that the parton luminosities are driven 
by different gluon PDFs of each group. These differences arise from the different methodologies 
and inputs (heavy-flavor treatment, values of $\alpha_s(M_Z)$, data selection, etc.) adopted by each 
PDF fitting group in their QCD analyses. The predictions using ABM11 are found to be generally lower with respect to the other sets, 
because the different methodology used in the ABM11 analysis leads to a lower gluon and 
smaller $\alpha_s(M_Z)$ value. A larger value of the predicted cross section using ABM11 
can be obtained by choosing a smaller value of the top-quark pole mass within the 
current uncertainties, see, e.g.,~\cite{Alekhin:2013nda}. 

Individual contributions to the total uncertainty on the prediction, arising from uncertainties of PDFs and
variation of $\alpha_s(M_Z)$, scales, and $m_t$, are studied
for each used PDF set separately. 
The results for the CT10NNLO PDFs are shown in Fig.~\ref{allPDFs-unc-pt-ct10}, while similar results obtained 
by using all the other PDF sets can be found in the Appendix.
The total uncertainty on the theory prediction for the total 
and absolute differential $t\bar{t}$ cross section is dominated by the PDF errors. At large $p^t_T$ and 
$|y^t|$, where regions of larger values of $x$ are probed, the PDF uncertainties 
increase since the gluon distribution in this range is poorly constrained at present. 
In the case of HERAPDF1.5 PDF, in addition to the experimental uncertainty, also parametrization and model 
uncertainties are estimated. The inclusion of the model uncertainty results in an increase of the cross 
section by approximately 20\%, related to the variation of the $q^2$ cut on the used DIS data. In the case of 
ABM11 PDFs, the $\alpha_s$ variation is already included in the quoted PDF error since in the ABM fit 
$\alpha_s$ is simultaneously determined together with the PDFs. 
\begin{figure}[h]
\begin{center}
\includegraphics[width=7.cm, angle=0]{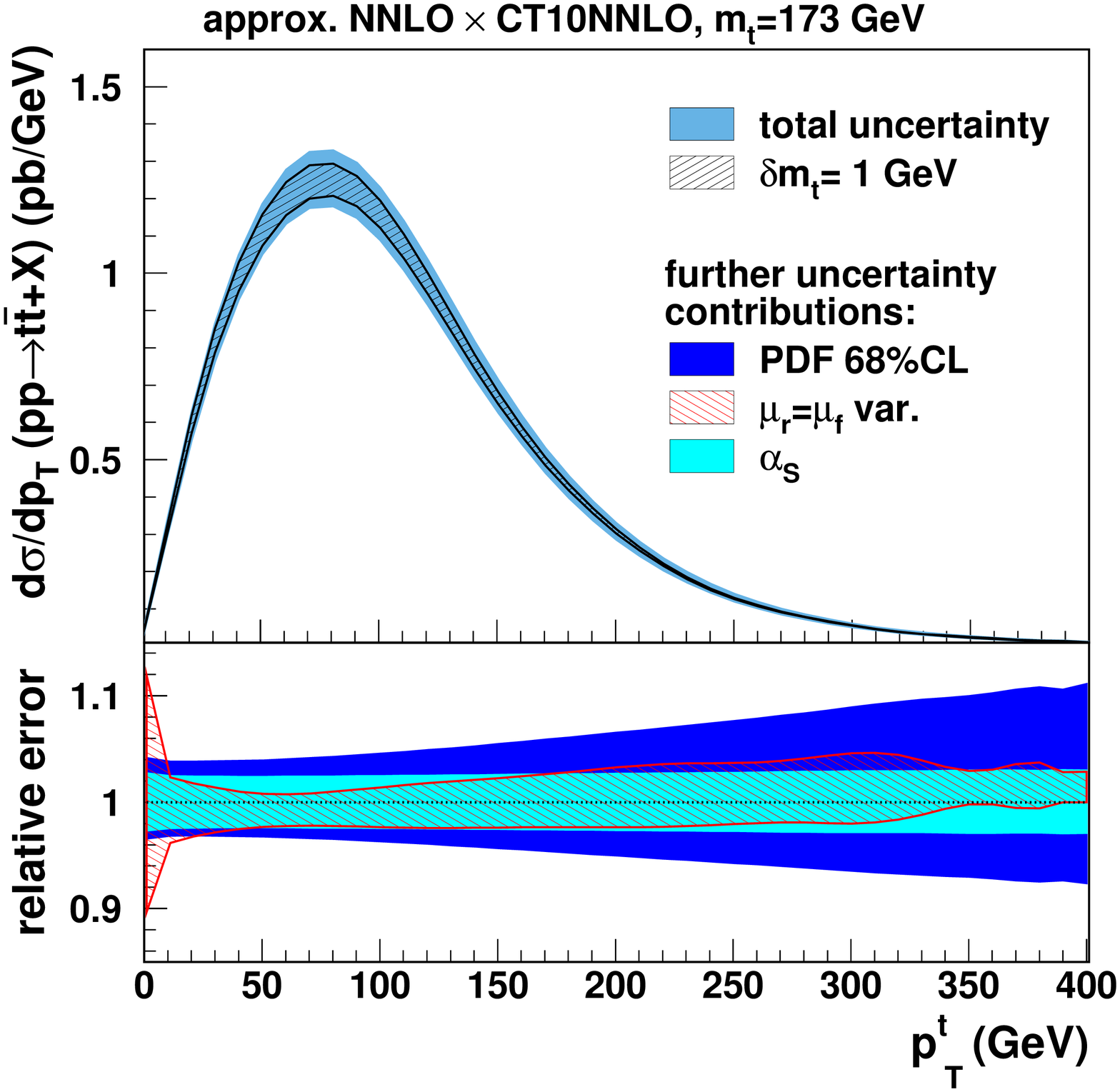}
\includegraphics[width=7.cm, angle=0]{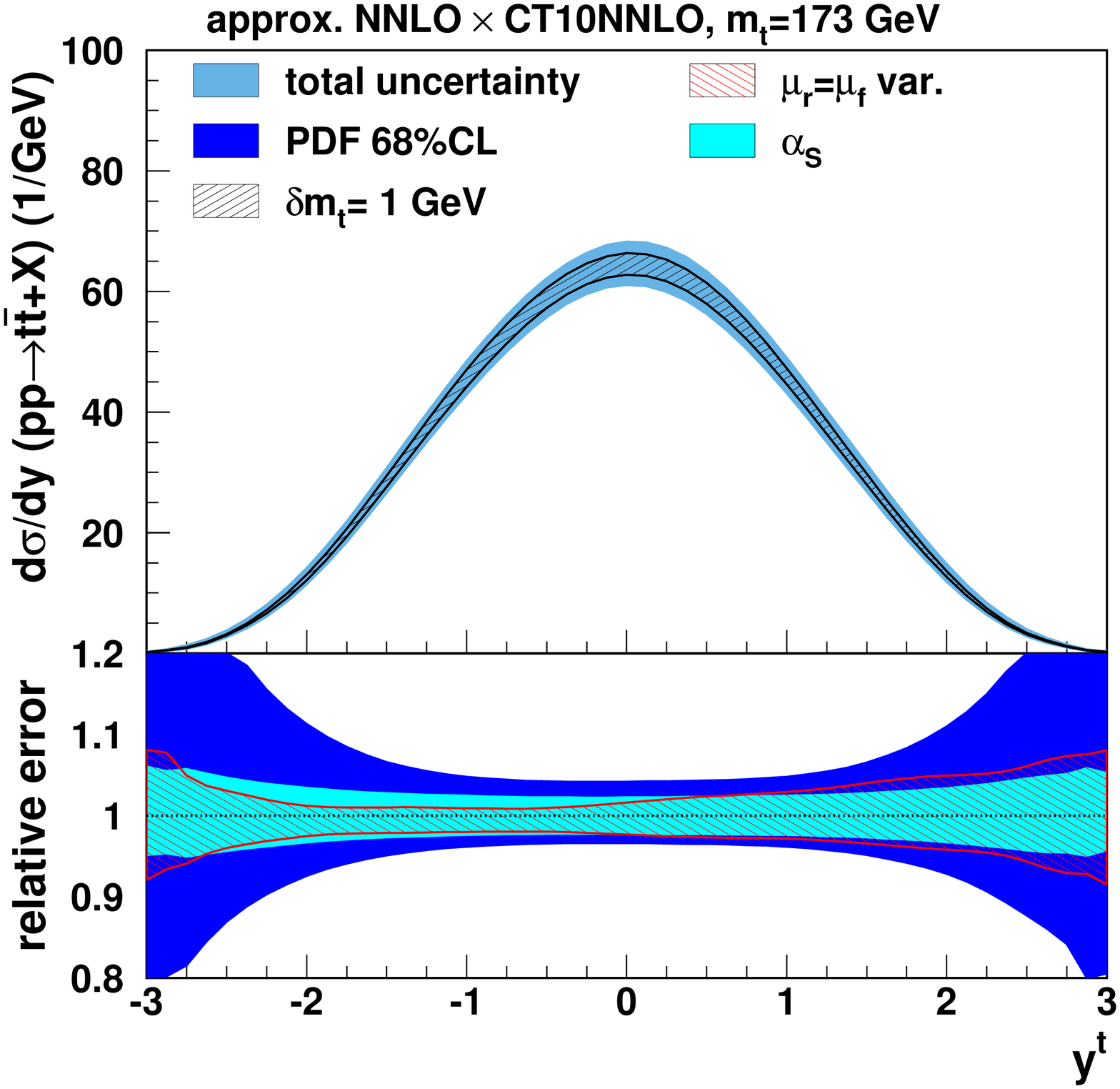}
\end{center}
\caption{The approximate NNLO predictions for $t\bar{t}$ production cross sections at the LHC at $\sqrt{S} = 7$ 
TeV as functions of $p^t_T$ (left) and $y^t$ (right) obtained by using CT10NNLO. The individual contributions of the uncertainties due 
to the PDFs (68\% CL), $\alpha_s(M_Z)$, scale, and $m_t$ variations, are shown separately by bands of different shades.
\label{allPDFs-unc-pt-ct10}}
\end{figure}

Variations of $m_t$ modify the magnitude and shape of the distribution at large-$p_T$ and 
represent the second dominant contribution to the total uncertainty. This reflects the strong sensitivity 
of the differential distributions to the top-quark mass. The usage of the $\overline{\textrm{MS}}$ scheme 
definition for the top-quark mass in the calculation is expected to improve the convergence of perturbation 
theory and, in turn, to reduce the scale dependence~\cite{Dowling:2013baa}. In some regions of the $p^t_T$ 
and $y^t$ spectra, the uncertainty associated to variations of $\alpha_s(M_Z)$ is sizable.
\begin{figure}[ht]
\includegraphics[width=5.25cm, angle=0]{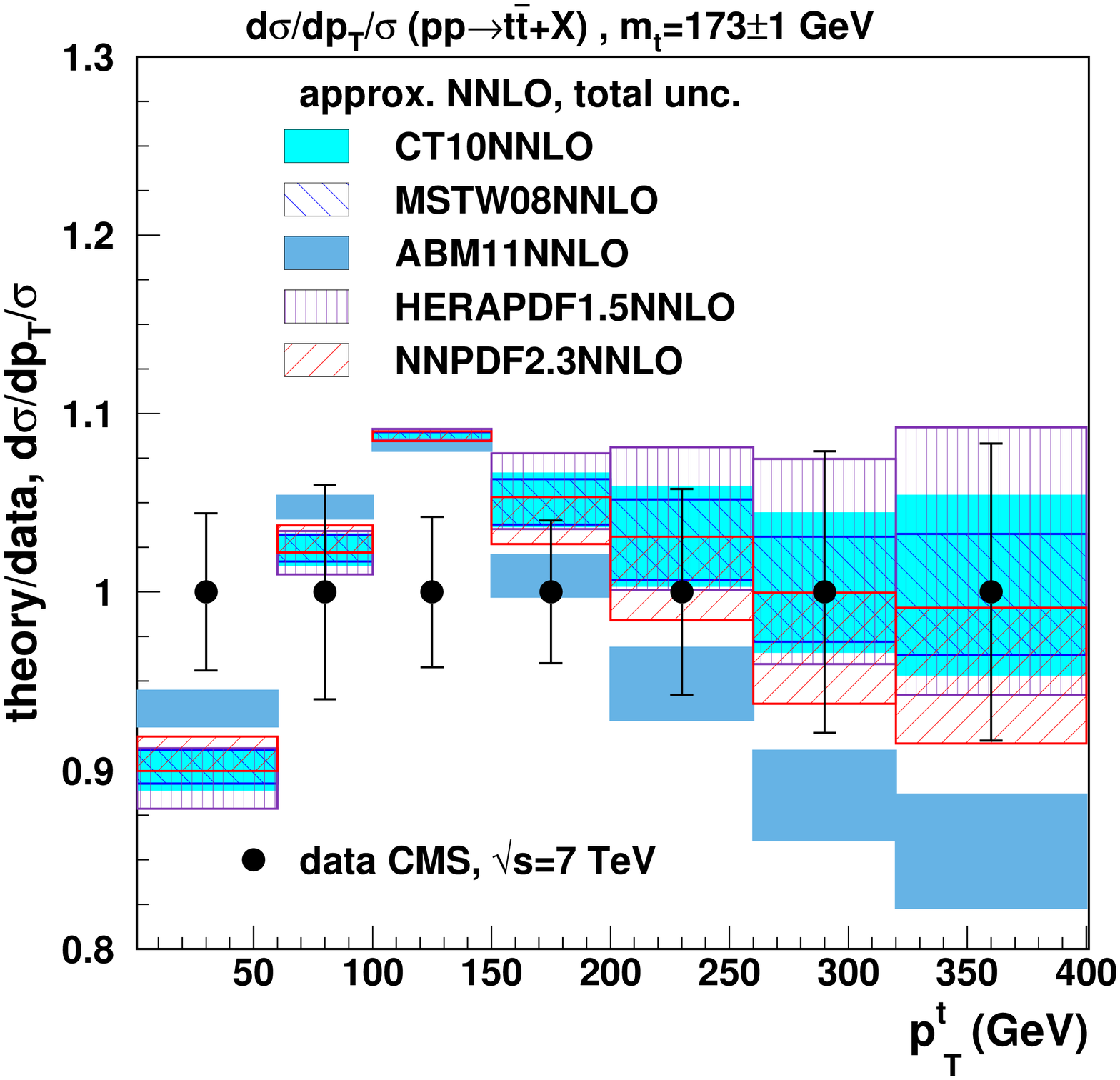}
\includegraphics[width=5.25cm, angle=0]{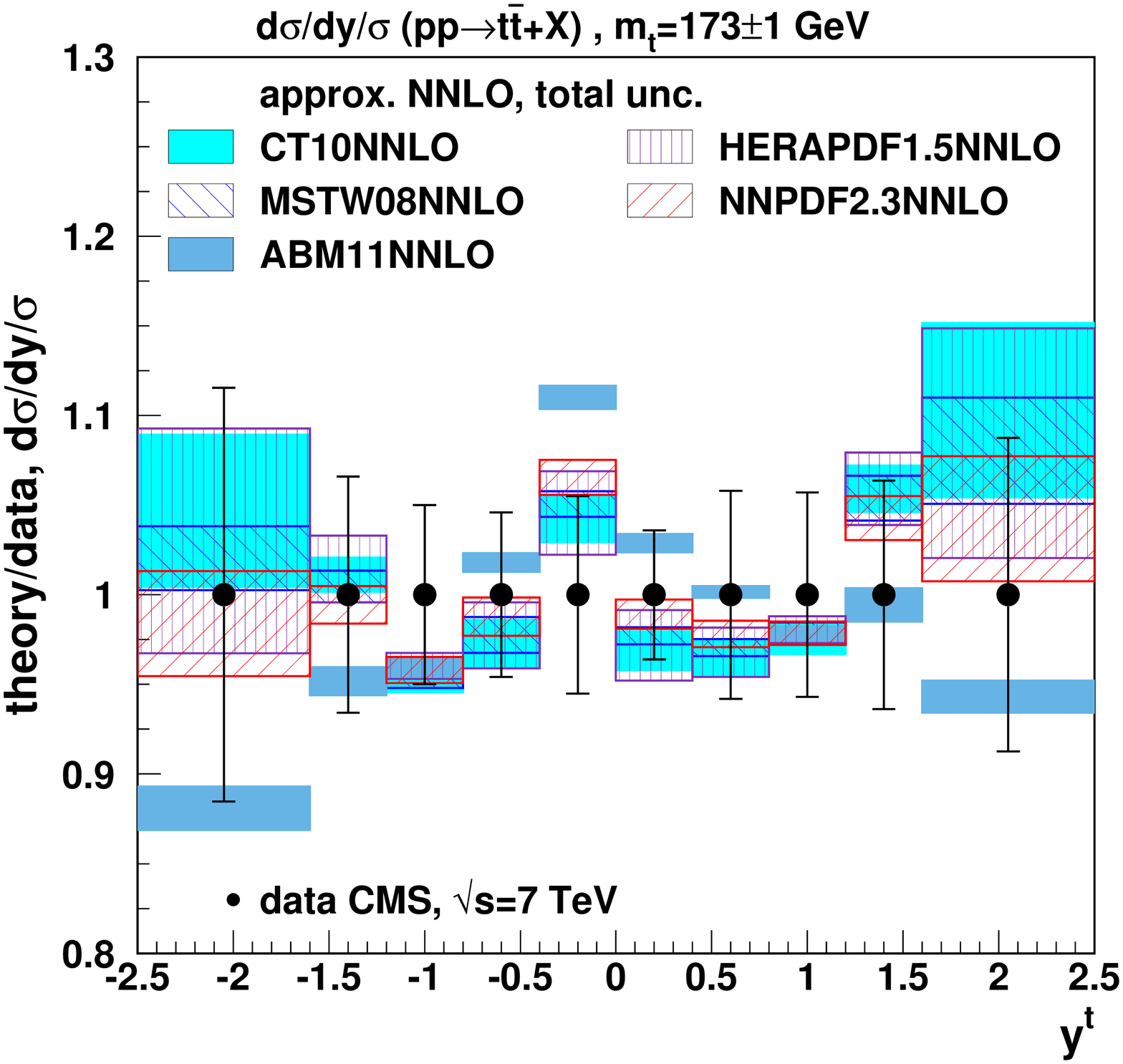}
\includegraphics[width=5.25cm, angle=0]{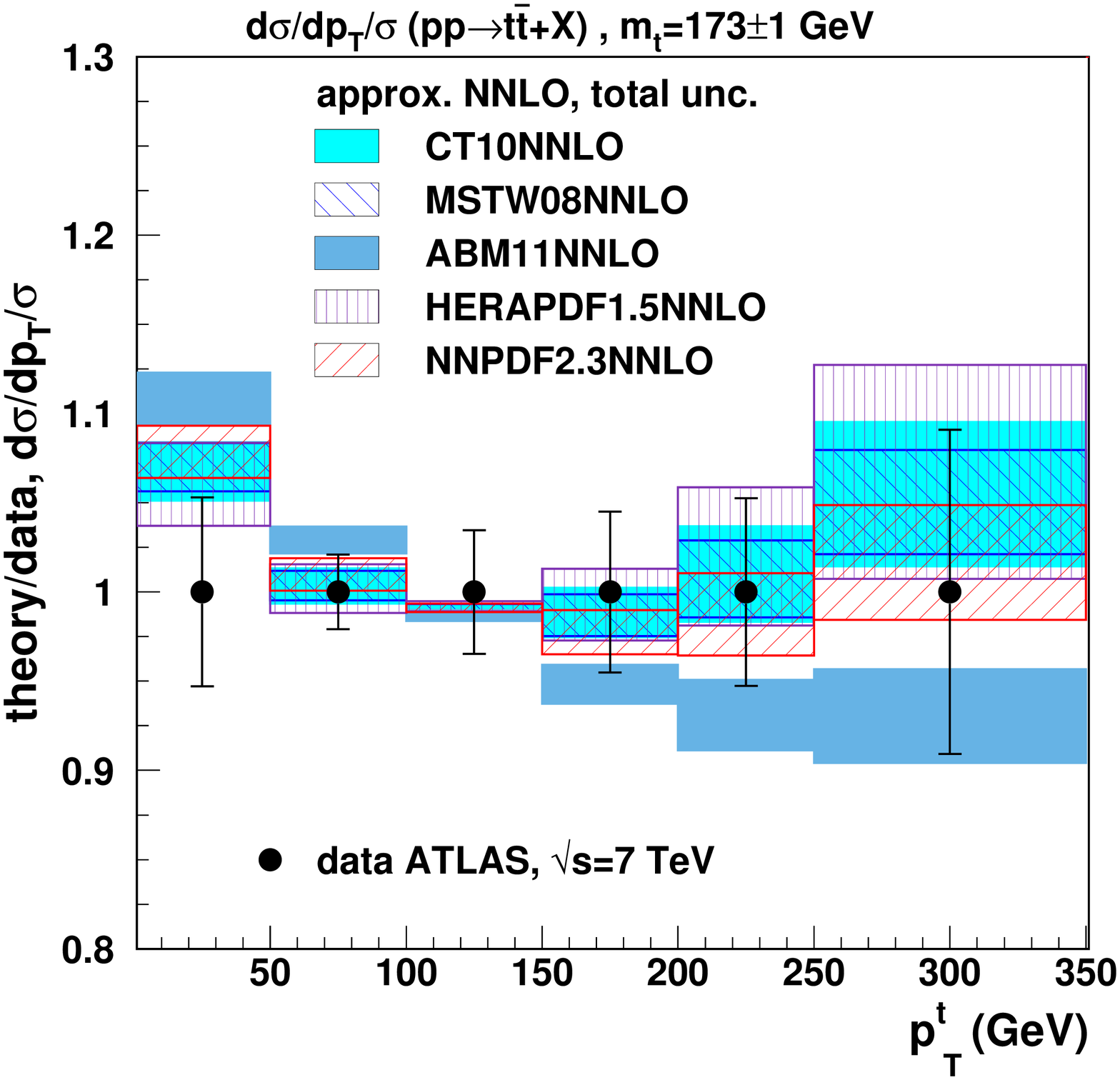}
\caption{The approximate NNLO predictions for $t\bar{t}$ production at the LHC at $\sqrt{S} = 7$ TeV, shown as 
functions of $p^t_T$ and $y^t$. The predictions with their total uncertainties, obtained by using different PDF 
sets (bands of different hatches), are presented as a ratio to the LHC measurements (filled symbols).
\label{allPDFs-unc-Th_dta}}
\end{figure}
The predictions for the normalized differential cross sections using different PDF sets are shown 
with their total uncertainties in Fig.~\ref{allPDFs-unc-Th_dta}, where these are compared to the recent measurements by 
the CMS and ATLAS collaborations at $\sqrt{S}=7$ TeV. The different contributions to the uncertainty of the 
theory prediction are studied individually and the results for CT10NNLO PDFs are illustrated in Fig.~\ref{CT10-unc}.
Similar results obtained by using other PDF sets are shown in the Appendix.

When theory predictions are compared to the data, CMS and ATLAS measurements exhibit some differences 
in the shape of the normalized $p^t_T$ distribution, in particular in the first and third bin.
In general, the first bins of transverse momentum $0\leq p^t_T\leq 200$ GeV and central bins
of rapidity $-1.5\leq y^t\leq 1.5$ have potentially more constraining power, because of smaller 
experimental uncertainties. At the present stage, even though the CMS and ATLAS measurements 
exhibit relatively large uncertainties, these data might have some impact in PDF determination once 
included in QCD fit analyses. 
\begin{figure}[*hb]
\includegraphics[width=5.25cm, angle=0]{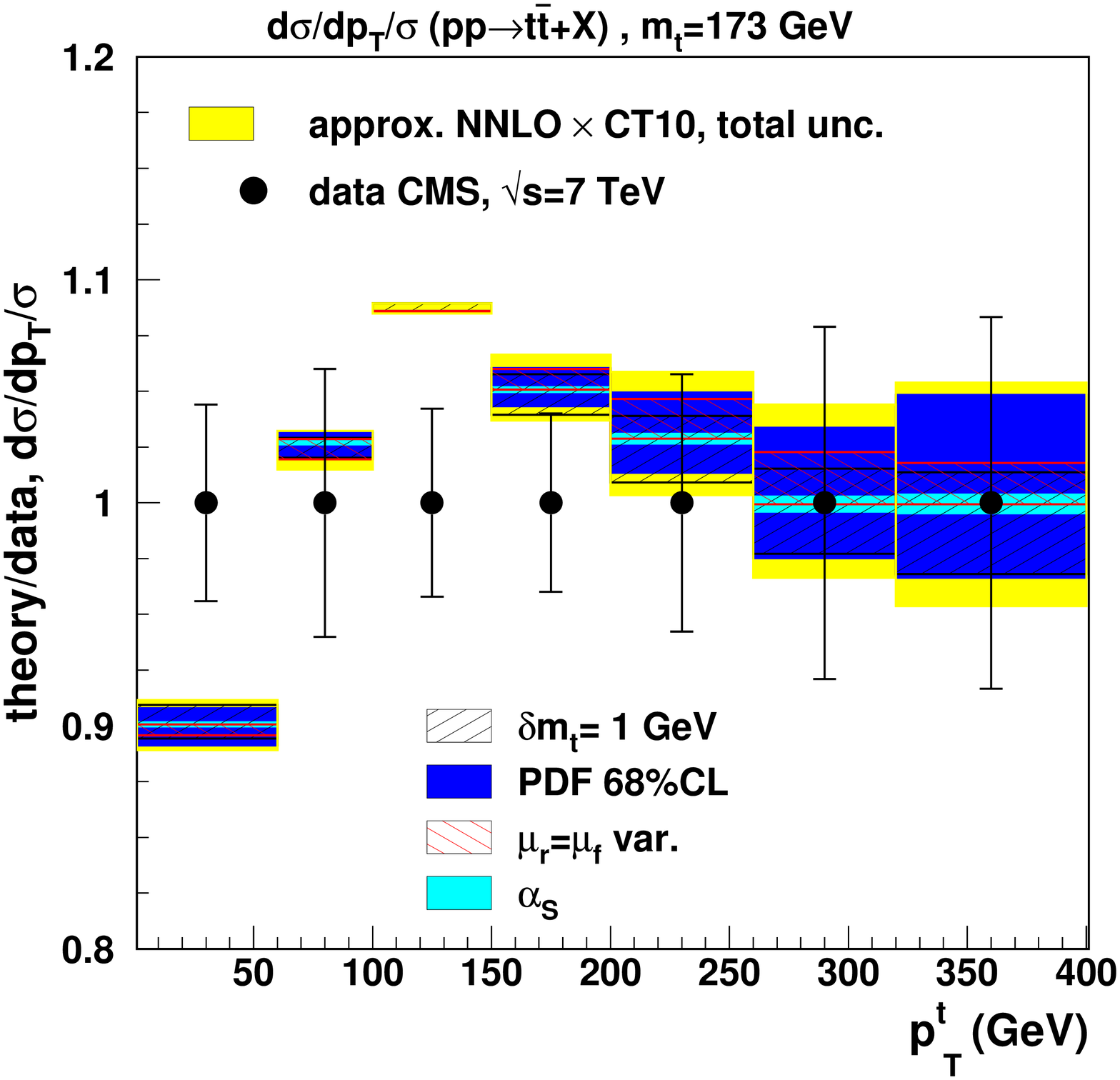}
\includegraphics[width=5.25cm, angle=0]{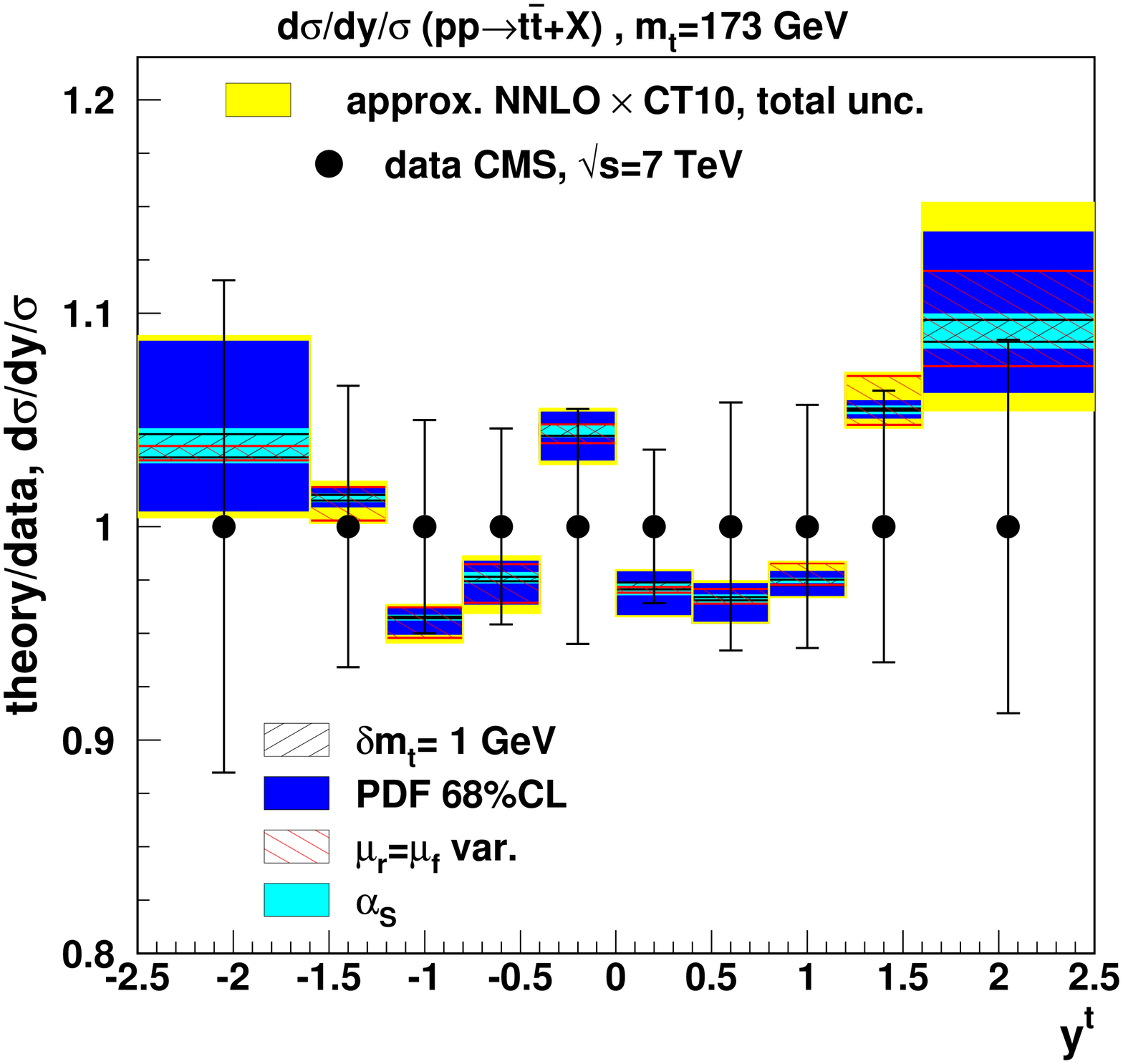}
\includegraphics[width=5.25cm, angle=0]{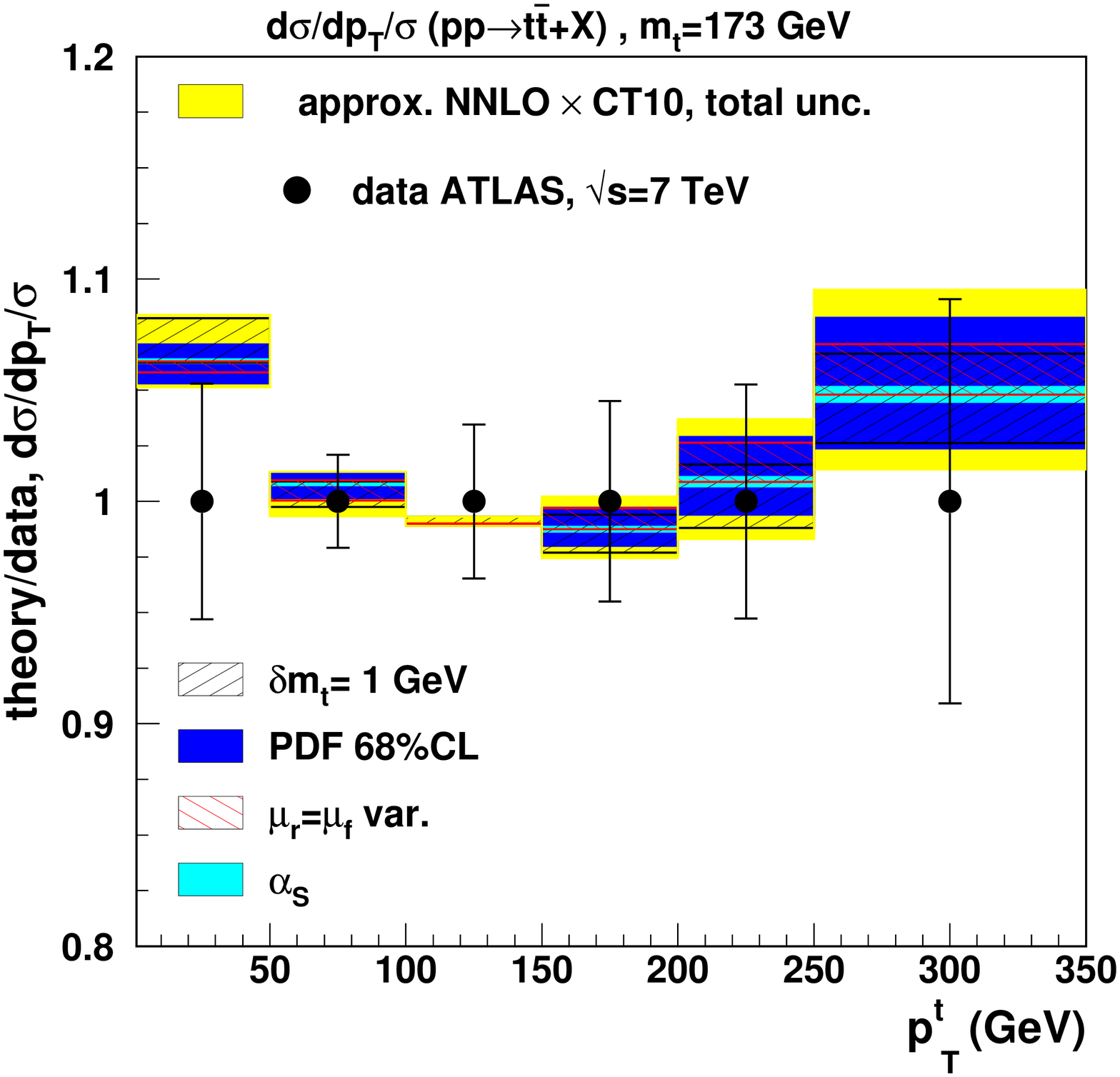}
\caption{The approximate NNLO prediction for $t\bar{t}$ production at the LHC at $\sqrt{S} = 7$ TeV obtained by 
using CT10NNLO PDF as a function of $p^t_T$ and $y^t$. The prediction is presented as a ratio to the LHC 
measurements (filled symbols). Individual contributions to the theoretical uncertainty are shown by the bands 
of different shades.\label{CT10-unc}}
\end{figure}
\clearpage

\subsection {QCD analysis using the $t\bar{t}$ production measurements.}

To illustrate the possible impact of the current available measurements of $t\bar{t}$ production on PDF determination,
we interfaced the \textsc{DiffTop} code to the \textsc{HERAFitter}~\cite{herafitter} platform for QCD analyses. 
Fast theoretical calculations are obtained by using grids generated by the \textsc{FastNLO} package~\cite{Kluge:2006xs,Wobisch:2011ij,Britzger:2012bs}. 
This allows the user to include measurements 
of differential $t\bar{t}$ production cross sections into NNLO QCD fits of PDFs.
The NNLO PDF fit performed here uses the parton evolution implemented in the \textsc{QCDNUM}~\cite{qcdnum} code,
that is the default parton-evolution package utilized in \textsc{HERAFitter}. 
The most important data sets for PDF determination are the combined HERA I measurements~\cite{hera1combined} 
of inclusive DIS, which we include in this analysis. 
Constraints on the $u$ and $d$-quark distributions in the $x$-range not properly covered by the HERA I measurements, 
are put by the CMS precise measurements of electron~\cite{CMS_e_asym} and muon~\cite{CMS_W_asym} charge asymmetry 
in $W$-boson production at $\sqrt{S}=7$ TeV. 
The theory predictions for the lepton charge asymmetry are obtained 
by using the $\textsc{MCFM}$ program at NLO and K-factors are applied in the NNLO fit. 

The $t\bar{t}$ measurements included in this analysis are the total inclusive cross sections at the 
LHC~\cite{ATLAS_tt_total_7,ATLAS_tt_total_8,CMS_tt_total_7, CMS_tt_total_8} at $\sqrt{S}$ of 7 and 8 TeV, 
and CDF~\cite{CDF_tt_total} Tevatron, as well as the normalized differential cross-sections~\cite{Chatrchyan:2012saa,Aad:2014zka} 
of $t\bar{t}$ production at the LHC at $\sqrt{S}=$ 7 TeV as a function of $p^t_T$. 
The procedure for the determination of the PDFs follows the approach used in the QCD fits of the
HERA~\cite{hera1combined} and CMS~\cite{CMS_W_asym} collaborations. 
In our PDF analysis, the TR'~\cite{Thorne:2006qt,Martin:2009ad} general mass variable flavor number scheme at NNLO is 
used for the treatment of heavy-quark contributions with input values of the heavy-quark 
masses given by $m_c$ = 1.4 GeV and $m_b$ = 4.75 GeV, while the choice of the QCD scales is $\mu_R$ = $\mu_F$ = $Q$.
The strong coupling constant is set to $\alpha_s (m_Z)$ = 0.1176 and the $Q^2$ range of the HERA data is 
restricted to the range $Q^2 \geq Q^2_{\textrm{min}}$ = 3.5 GeV$^2$.
The following independent combination of parton distributions is chosen at the QCD evolution initial scale $Q_0^2 = 1.9$ GeV$^2$ 
in the fitting procedure: $xu_{\textrm{v}}(x)$, 
$xd_{\textrm{v}}(x)$, $xg(x)$ and $x\overline{\textrm{U}}(x)$, $x\overline{\textrm{D}}(x)$, where
$x\overline{\textrm{U}}(x) = x\bar{u}(x)$ and $x\overline{\textrm{D}}(x) = x\bar{d}(x) + x\bar{s}(x)$. 
At the scale $Q_0$, the parton distributions are represented by
\begin{eqnarray}
x u_\textrm{v}(x) &=& A_{u_{\textrm{v}}} ~  x^{B_{u_{\textrm{v}}}} ~ (1-x)^{C_{u_{\textrm{v}}}} ~(1+D_{u_{\textrm{v}}} x+E_{u_{\textrm{v}}} x^2) ,
\label{eq:uv}\\
x d_\textrm{v}(x) &=& A_{d_{\textrm{v}}} ~ x^{B_{d_{\textrm{v}}}} ~ (1-x)^{C_{d_{\textrm{v}}}},  
\label{eq:dv}\\
x \overline {\textrm{U}}(x) &=& A_{\overline {\textrm{U}}} ~ x^{B_{\overline {\textrm{U}}}} ~ (1-x)^{C_{\overline {\textrm{U}}}},
\label{eq:Ubar}\\
x \overline {\textrm{D}}(x) &=& A_{\overline{\textrm{D}}} ~ x^{B_{\overline{\textrm{D}}}} ~ (1-x)^{C_{\overline{\textrm{D}}}}, 
\label{eq:Dbar}\\
x g(x) &=& A_{g} ~ x^{B_{g}} ~ (1-x)^{C_{g}} 
+ A'_{g} ~ x^{B'_{g}} ~ (1-x)^{C'_{g}}.  
\label{eq:g}
\end{eqnarray}
The normalization parameters $A_{u_{\textrm{v}}}$, $A_{d_\textrm{v}}$, $A_g$ are determined by the QCD sum 
rules, the $B$ parameter is responsible for small-$x$ behavior of the PDFs, and the parameter $C$ 
describes the shape of the distribution as $x \to 1$. A flexible form for the gluon distribution is 
adopted with the choice of $C'_g=25$ motivated by the approach of the MSTW group~\cite{Thorne:2006qt,Martin:2009ad}.
The analysis is performed by fitting 14 free parameters in Eqs.~(\ref{eq:uv}-\ref{eq:g}). 
Additional constraints $B_{\overline{\textrm{U}}} = B_{\overline{\textrm{D}}}$ and $A_{\overline{\textrm{U}}} = A_{\overline{\textrm{D}}}(1 - f_s)$ 
are imposed by $f_s$ that is the strangeness fraction defined as $f_s = \bar{s}/( \bar{d} + \bar{s})$, which is fixed 
to $f_s=0.31\pm0.08$ as in the analysis of Ref.~\cite{Martin:2009ad}. 

A comparison of the PDFs resulting from the fit obtained by using only the HERA DIS data 
and that obtained by employing the HERA DIS in addition to the CMS lepton charge asymmetry measurements, 
shows effects on the central value of the light-quark distributions and on the reduction of the uncertainties 
which are similar to the findings reported by the CMS collaboration in a recent QCD analysis~\cite{CMS_W_asym} at NLO.
A slight reduction of the uncertainty of the gluon distribution in the HERA DIS + CMS lepton asymmetry fit 
with respect to the fit including the HERA DIS only, is ascribed to the improved constraints on the light-quark 
distributions through the sum rules. 

\begin{figure}[h]
\includegraphics[width=8.cm, angle=0]{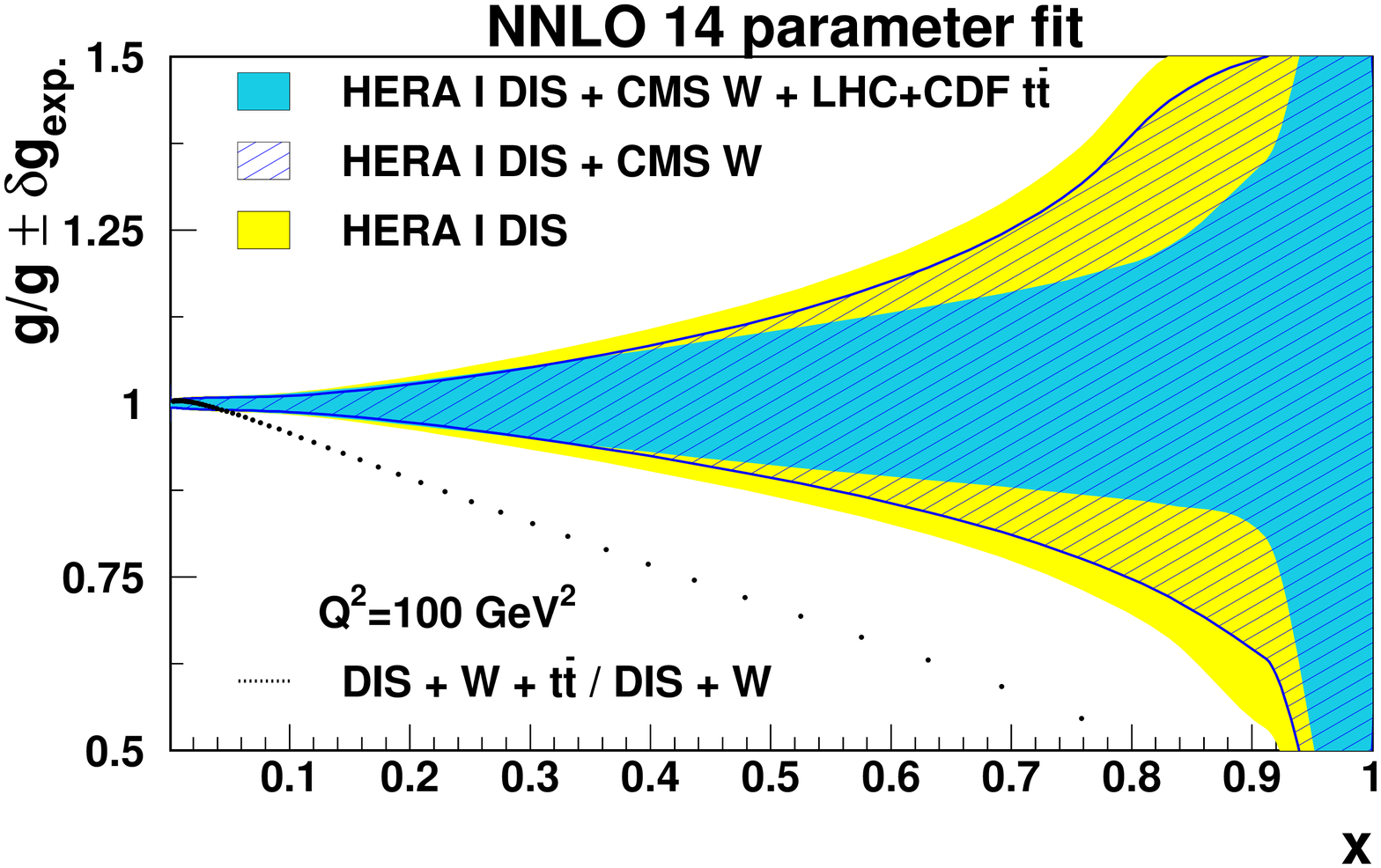}
\includegraphics[width=8.cm, angle=0]{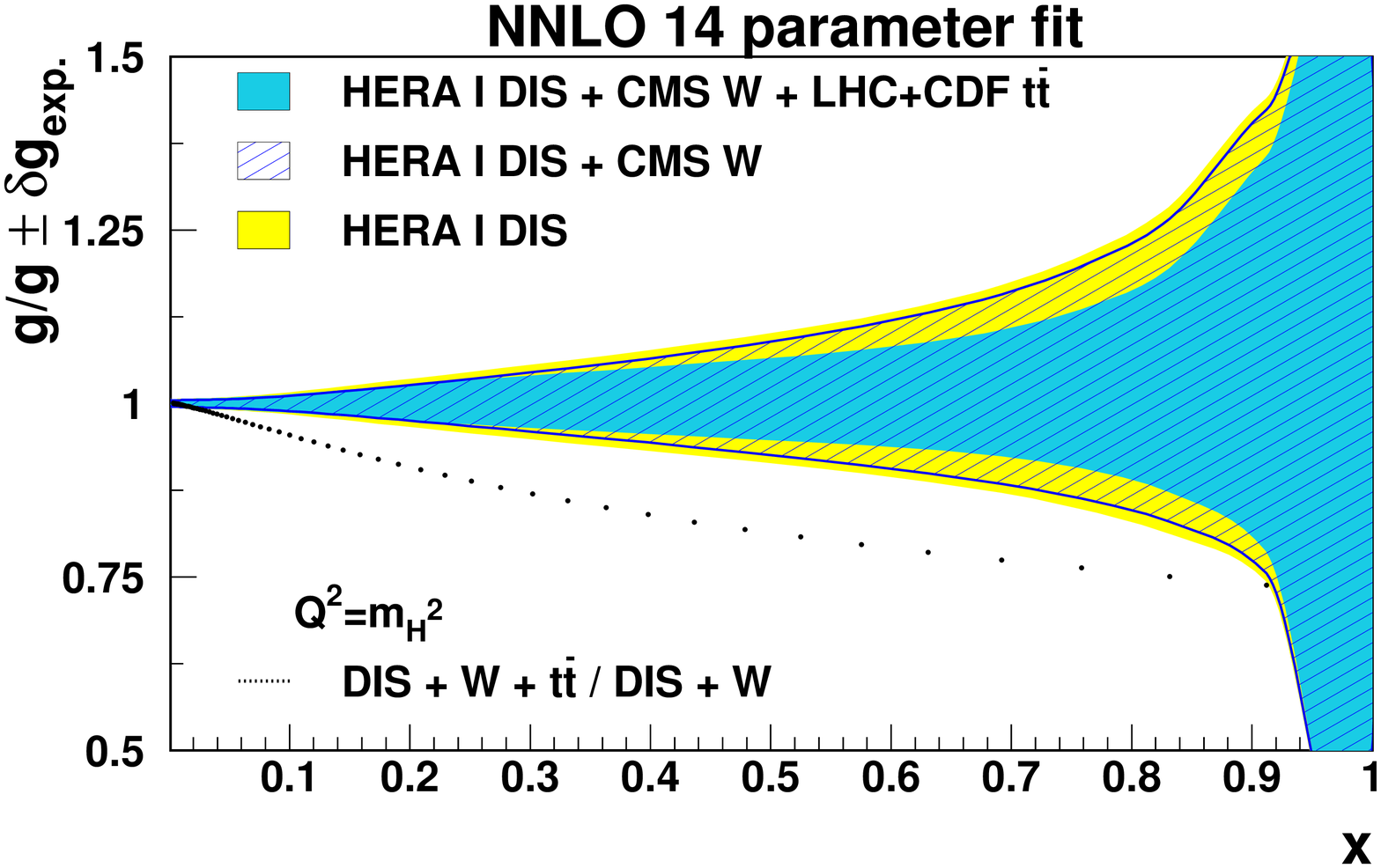}
\caption{Uncertainties of the gluon distribution as a function of $x$, as obtained in our NNLO fit 
by using: inclusive DIS measurements only (light shaded band), DIS and $W$ lepton charge asymmetry 
data (hatched band), and DIS, lepton charge asymmetry and the $t\bar{t}$ production measurements 
(dark shaded band), shown at the scales of $Q^2=100$ GeV$^2$ (left) and $Q^2=m_H^2$ (right). 
The ratio of $g(x)$ obtained in the fit including $t\bar{t}$ data to that 
obtained by using DIS and lepton charge asymmetry, is represented by a dotted line.   
\label{fig:hera_vsall}}
\end{figure}

The inclusion the $t\bar{t}$ measurements in the current NNLO PDF fit leads to a change 
of the shape of the gluon distribution and a moderate improvement of its uncertainty 
at large $x$. This is observed in particular at high scales, as illustrated in Fig.~\ref{fig:hera_vsall}. 
By increasing the scale, the quantitative reduction of the uncertainty of the gluon 
distribution remains similar, but it sets in at lower values of $x$.
A similar effect is observed, although less pronounced, 
when only the total or only the differential $t\bar{t}$ 
cross section measurements are included in the fit. The results of our full PDF fit, demonstrating a 
moderate improvement of the uncertainty 
on the gluon distribution, confirm the observation reported in the reweighting analysis~\cite{Czakon:2013tha} 
which uses only the total $t\bar{t}$ cross sections.

The analysis presented here uses the normalized differential cross sections for $t\bar{t}$ production. 
The use of normalized data leads to partial cancellation or reduction of the experimental uncertainties. 
However, a significant amount of information is lost by normalizing the data, in 
particular, in connection to uncertainty correlations, which are currently not provided by the experimental collaborations.
Measurements of absolute differential cross sections supplied by full information about correlation of 
experimental uncertainties are of crucial importance to fully exploit the potential of the $t\bar{t}$ 
production to constrain the gluon distribution. Furthermore, the dependence of the experimentally measured 
differential $t\bar{t}$ cross section on the assumptions on the top-quark mass used in the Monte Carlo simulations 
used for efficiency calculation, is necessary. 
In the future, a reduction of the statistic and systematic uncertainties in the high-energy run of the LHC will 
be of clear advantage. A simultaneous determination of the gluon distribution and the top-quark mass is of 
particular interest.

\section{Summary and conclusions}

In this paper, we present results for the differential cross section of $t\bar{t}$ production 
at approximate NNLO that are of interest for phenomenological studies at hadron colliders. 
This calculation is implemented into the flexible computer code \textsc{DiffTop}, which is a useful tool   
for precision studies in QCD and phenomenological applications of the $t\bar{t}$ differential cross section.

In particular, details of the predicted distributions of top-quark transverse momentum and rapidity, 
generated by using different PDF sets are studied and compared to the recent measurements 
by the CMS and ATLAS collaborations. Individual uncertainties due to variations of PDFs, scale, 
$\alpha_s$ and $m_t$, are analyzed.
The \textsc{DiffTop} code has been interfaced to \textsc{FastNLO} for fast evaluations 
of the theory predictions, and it has been included into the \textsc{HERAFitter} framework for 
QCD analysis to determine the PDFs of the proton. 
For the first time, it is possible to include measurements 
of the differential cross sections of $t\bar{t}$ production at the LHC into 
a full PDF fit. 
We studied the impact of the recent measurements of the inclusive and differential $t\bar{t}$ 
production cross sections in a PDF fit at NNLO together with the HERA I inclusive DIS data and the
CMS measurements of the lepton charge asymmetries in $W$ boson production. 
Given the current experimental precision of $t\bar{t}$ measurements, 
a moderate improvement of the uncertainty on the gluon distribution at high $x$ 
is observed once the total and differential $t\bar{t}$ cross sections are included in the QCD analysis. 
Measurements of the $p^t_T$ and $y^t$ differential cross 
sections of $t\bar{t}$ production in hadron collisions 
are going to play an important role in constraining the gluon PDF 
at large-$x$ as the LHC will reach higher experimental precision in the forthcoming run II. 
Given the correlations between the strong coupling $\alpha_s(M_Z)$, top-quark mass and gluon PDF, these 
measurements could be used to constrain the large-$x$ gluon distribution and the top-quark mass simultaneously. 
In particular, investigations of absolute differential cross sections will bring complementary information 
related to the magnitude and other details of the distributions, which will be crucial to improve the 
constraining power of the experimental data.

\section*{Acknowledgments}

We would like to thank Ben Pecjak for providing mathematica files 
with expressions to cross-check part of the results on the matching conditions.
We would also like to thank Nick Kidonakis, Maria Aldaya, Ringaile Placakyte and Aleko Khukhunaishvili 
for useful discussions.
We are grateful to Daniel Britzger (\textsc{FastNLO} Collaboration) for assistance 
in the implementation of \textsc{DiffTop} in \textsc{FastNLO}, 
and for providing us with the corresponding tables.
This work is supported in part by the 
``Initiative and Networking Fund of the Helmholtz Association (HGF) under the contract S0-072'',
by Deutsche Forschungsgemeinschaft in Sonderforschungs\-be\-reich/Transregio~9, 
by the European Commission through contract PITN-GA-2010-264564 ({\it LHCPhenoNet}), 
and it was realized within the scope of the PROSA collaboration.

\vspace{0.5cm}
{\bf Note Added.}
During the review process of this paper, very preliminary results for the exact calculation 
for the differential cross sections for $t\bar{t}$ production at NNLO in QCD
were presented at the {\it TOP2014} conference in Cannes, France.
Details can be found at~\cite{exact-preliminar}.

\bibliographystyle{h-elsevier3}

%\bibliography{top-draft}
\newpage
\section*{Appendix} 
\begin{figure}[h!]
\begin{center}
\includegraphics[width=5.5cm, angle=0]{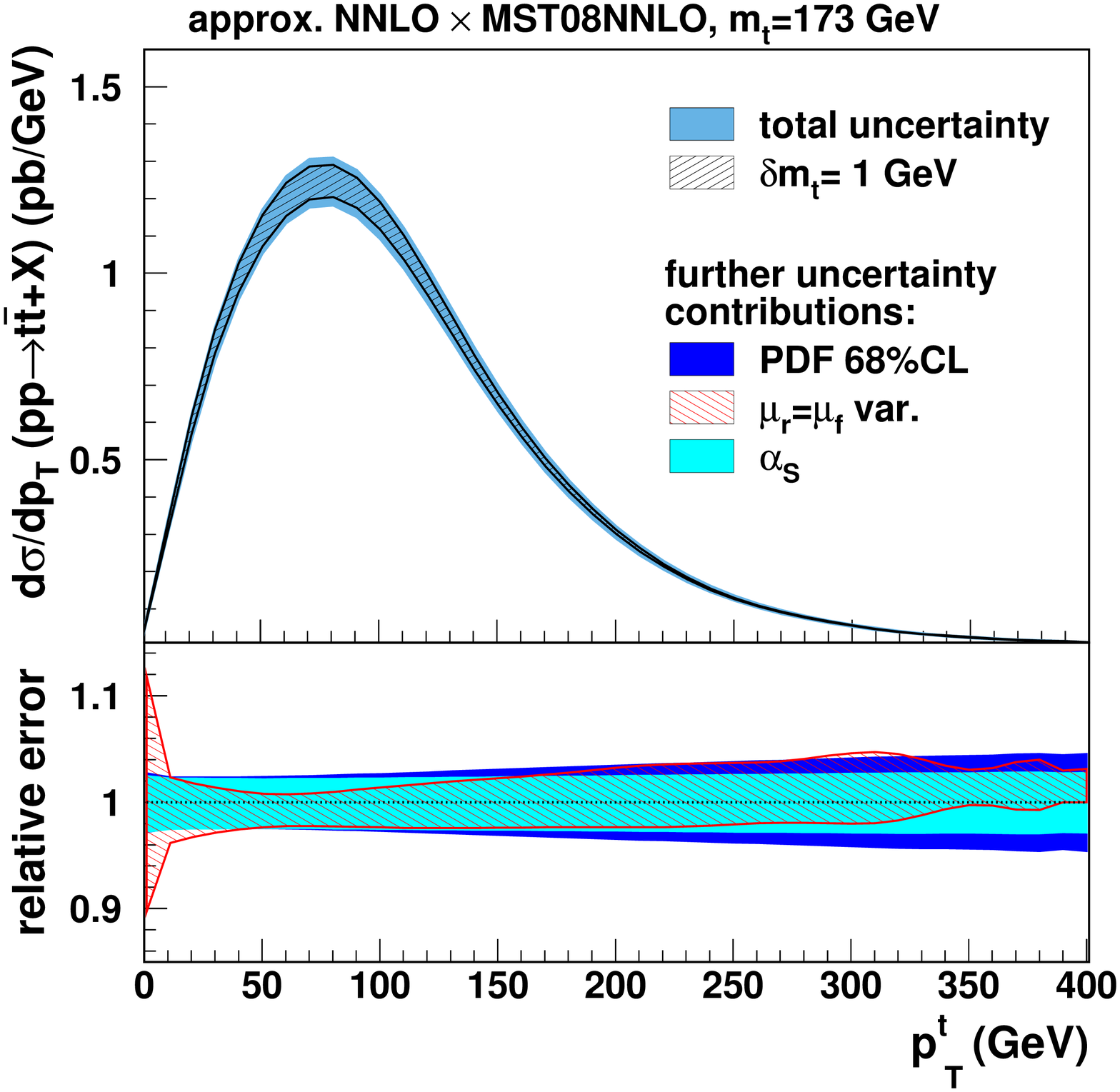}
\includegraphics[width=5.5cm, angle=0]{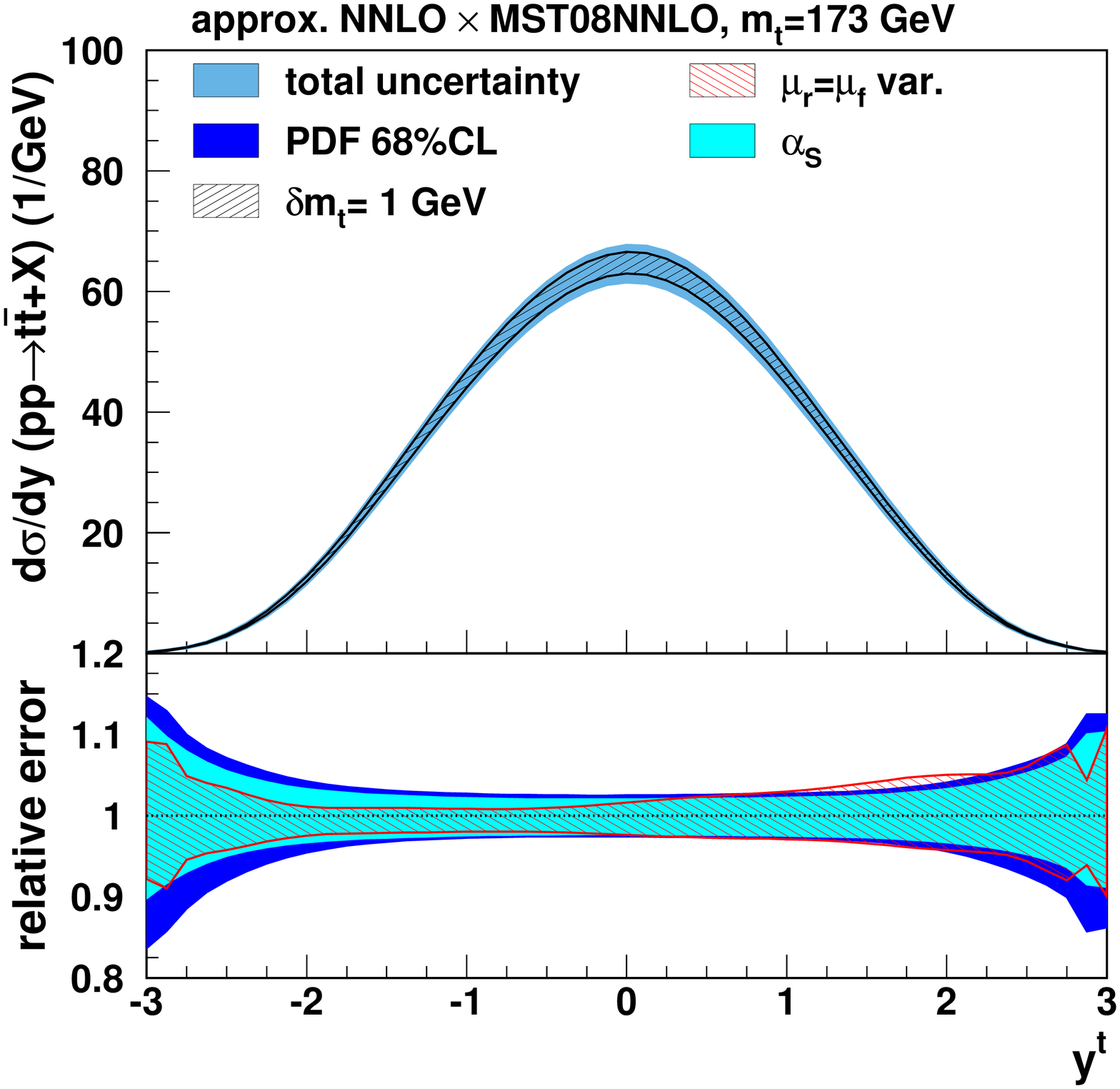}\\
\includegraphics[width=5.5cm, angle=0]{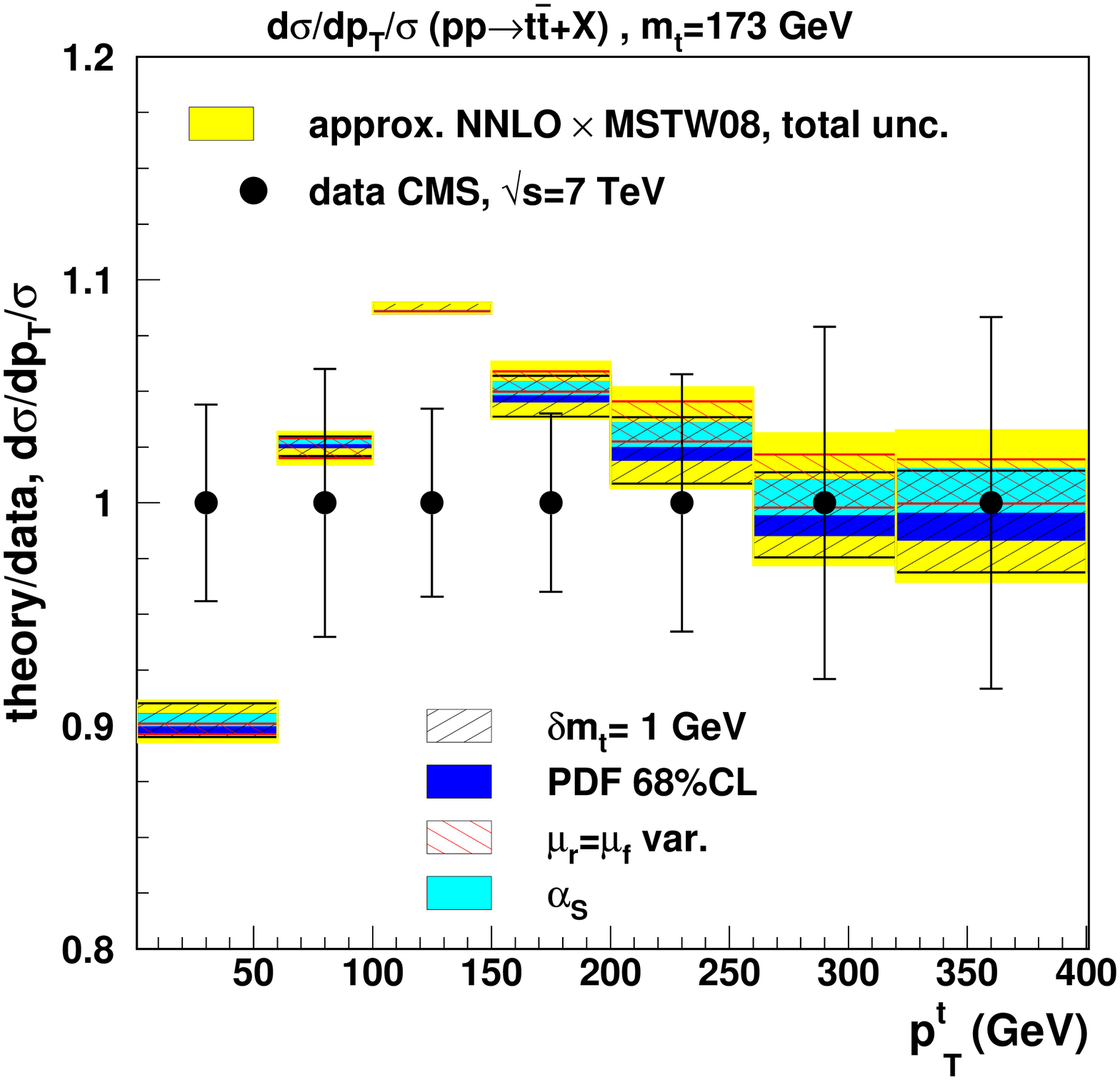}
\includegraphics[width=5.5cm, angle=0]{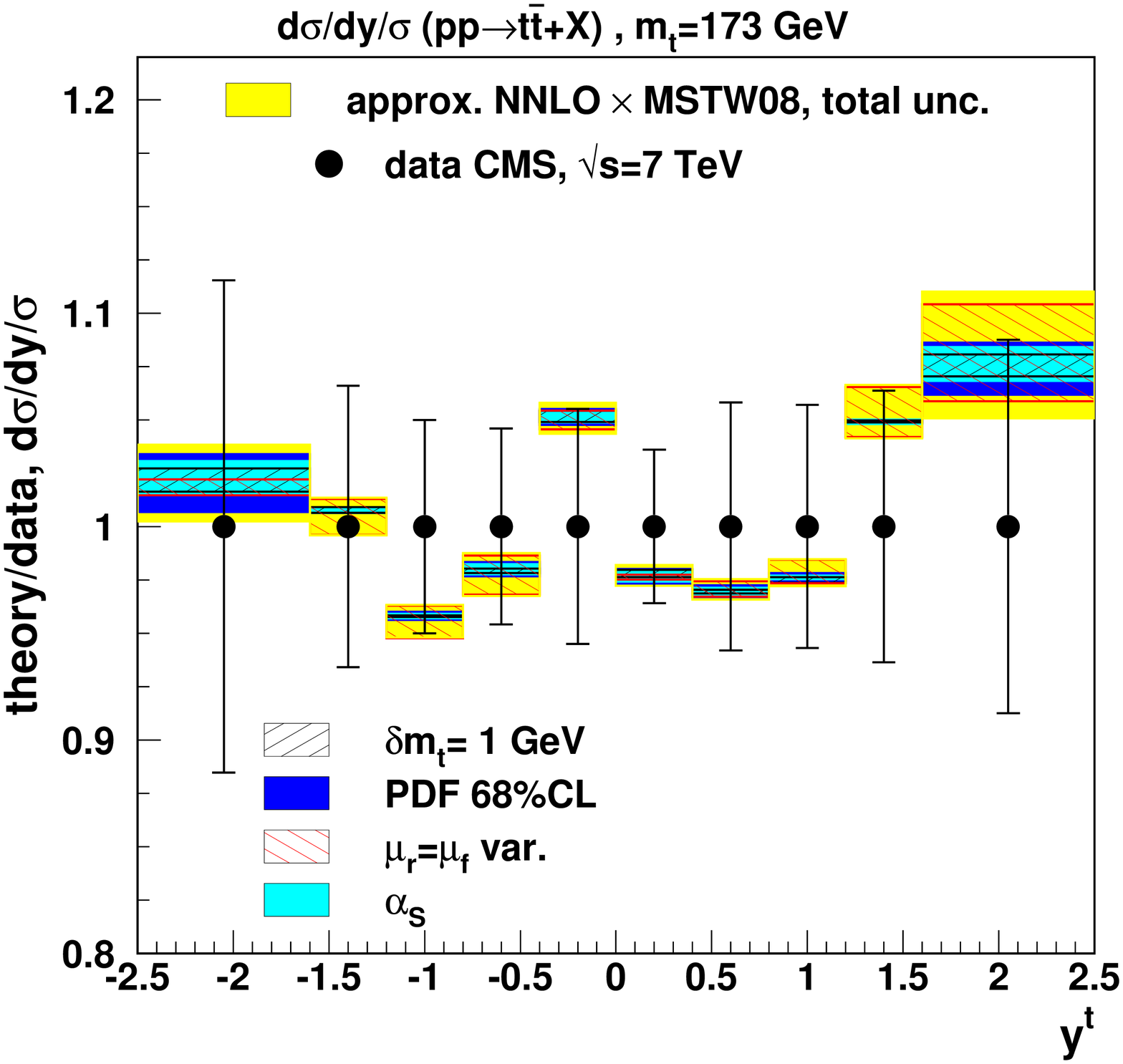}\\
\includegraphics[width=5.5cm, angle=0]{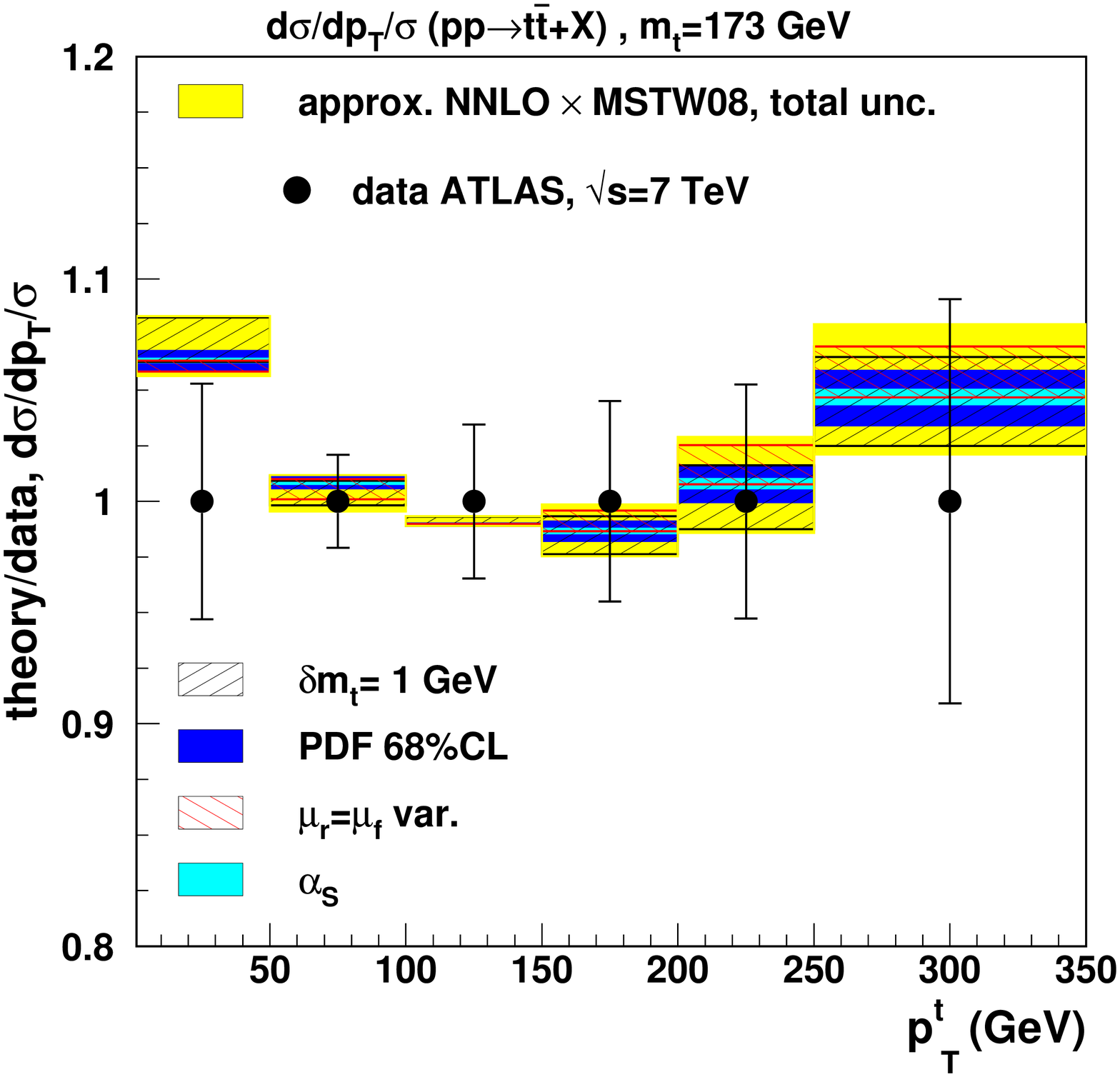}
\end{center}
\caption{\small Upper inset: approximate NNLO predictions and the relative uncertainty for $t\bar{t}$ production cross 
sections as a function of $p^t_T$ and $y^t$ obtained by using MSTW08NNLO. Individual contributions of 
uncertainties due to PDF (68\% CL), $\alpha_s(M_Z)$, scale and $m_t$ variations are shown by bands of 
different shades. Lower inset: Ratio of theory over data (light shaded band) for $p^t_T$ and $y^t$ as 
compared to the LHC measurements (filled circles).
\label{mstw08-unc}}
\end{figure}
\begin{figure}
\begin{center}
\includegraphics[width=5.5cm, angle=0]{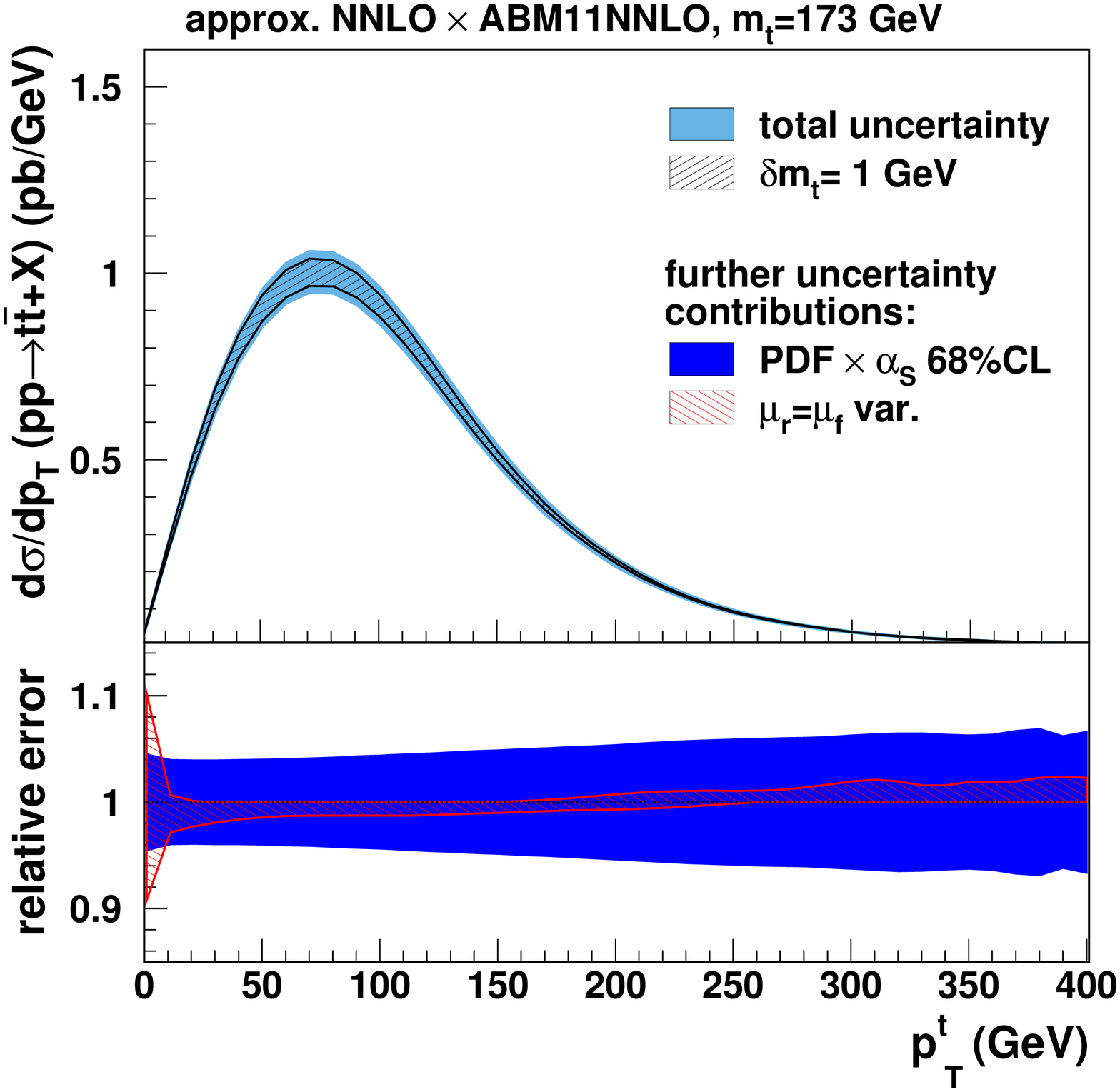}
\includegraphics[width=5.5cm, angle=0]{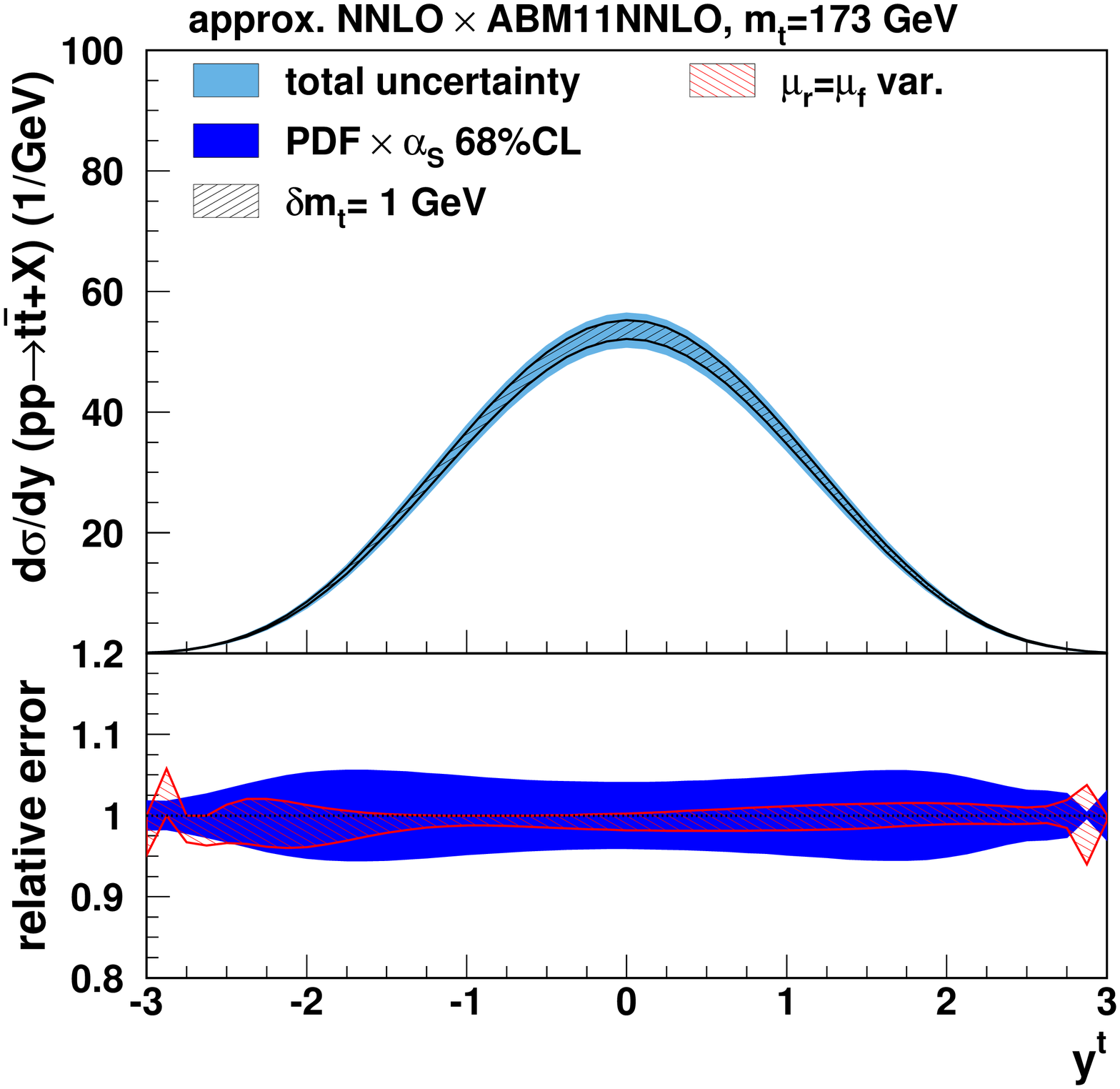}
\\
\includegraphics[width=5.5cm, angle=0]{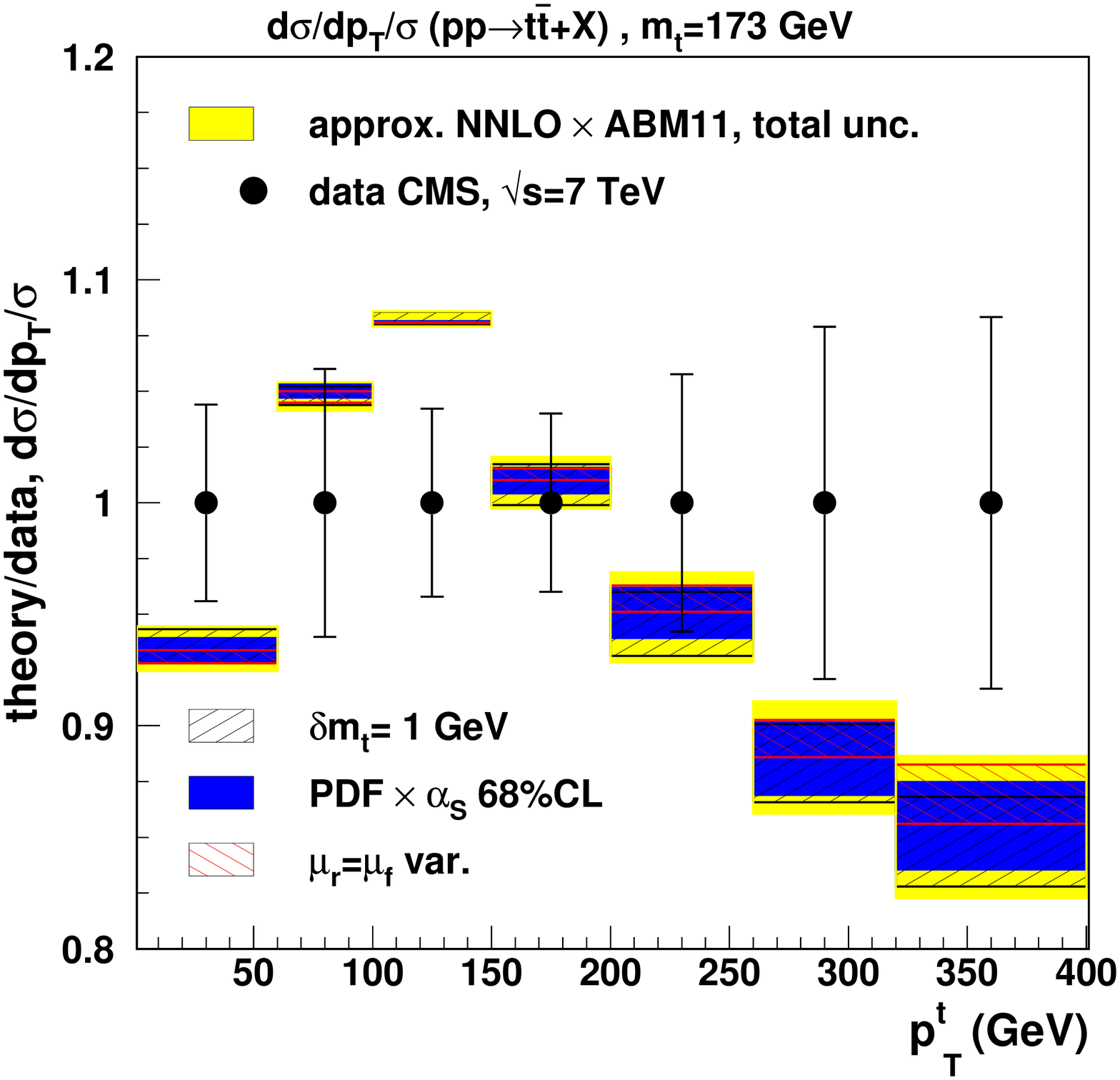}
\includegraphics[width=5.5cm, angle=0]{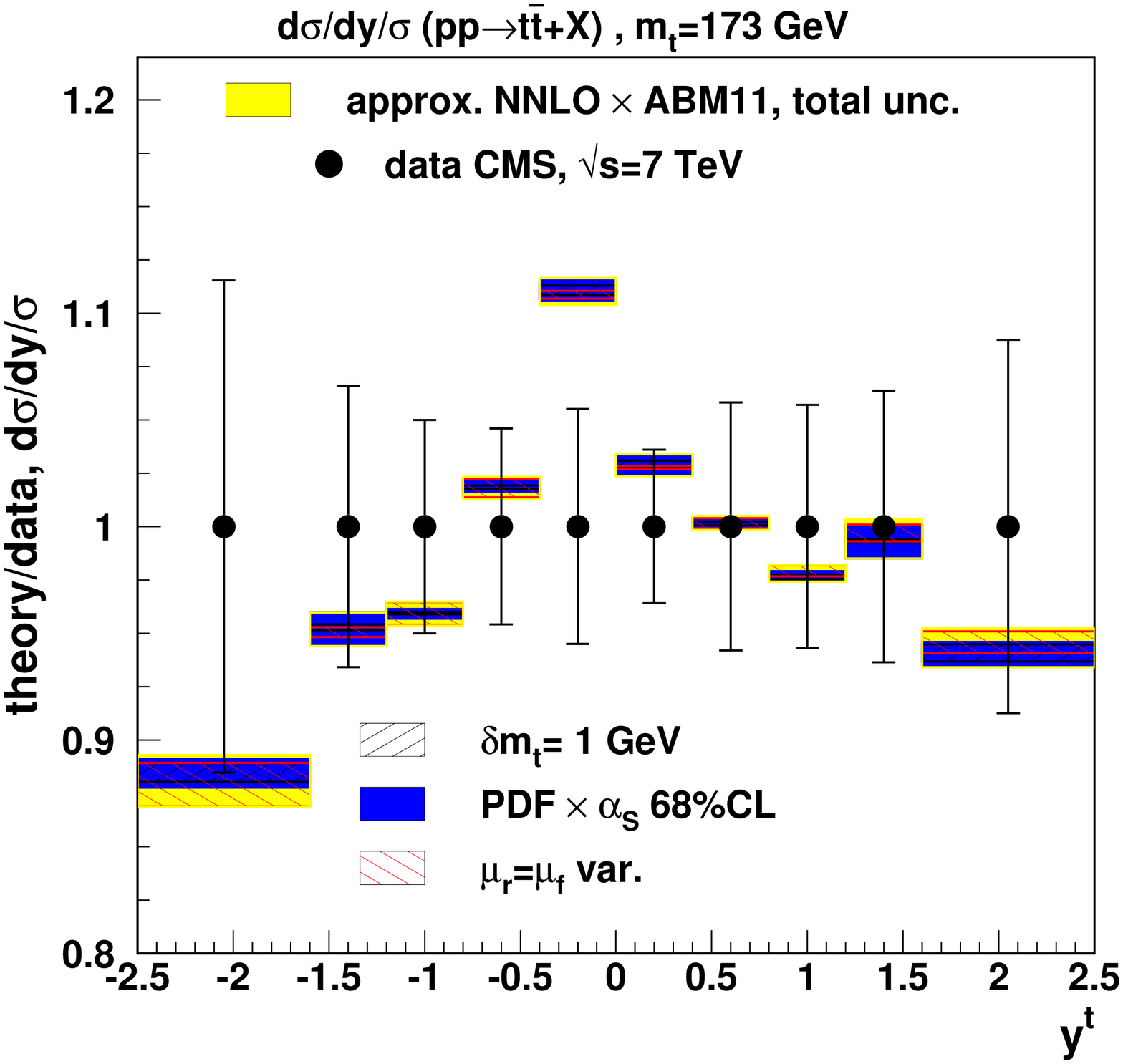}
\\
\includegraphics[width=5.5cm, angle=0]{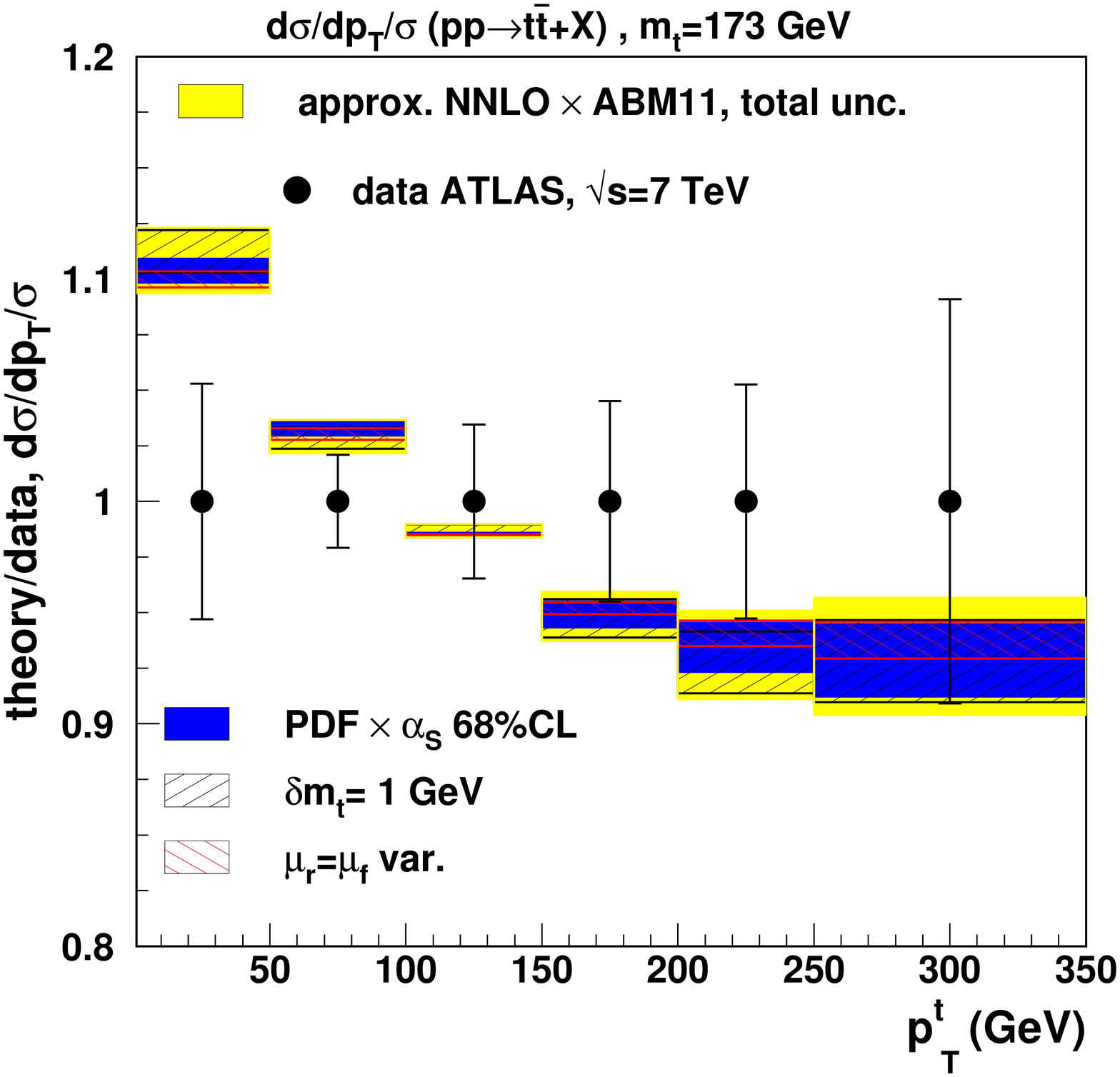}
\end{center}
\caption{Same as in Fig.~\ref{mstw08-unc} using ABM11 PDFs. 
Here, the uncertainty on $\alpha_s(M_Z)$ is included into the PDF uncertainty, see description in the text.
\label{abm11-unc}}
\end{figure}
\begin{figure}
\begin{center}
\includegraphics[width=5.5cm, angle=0]{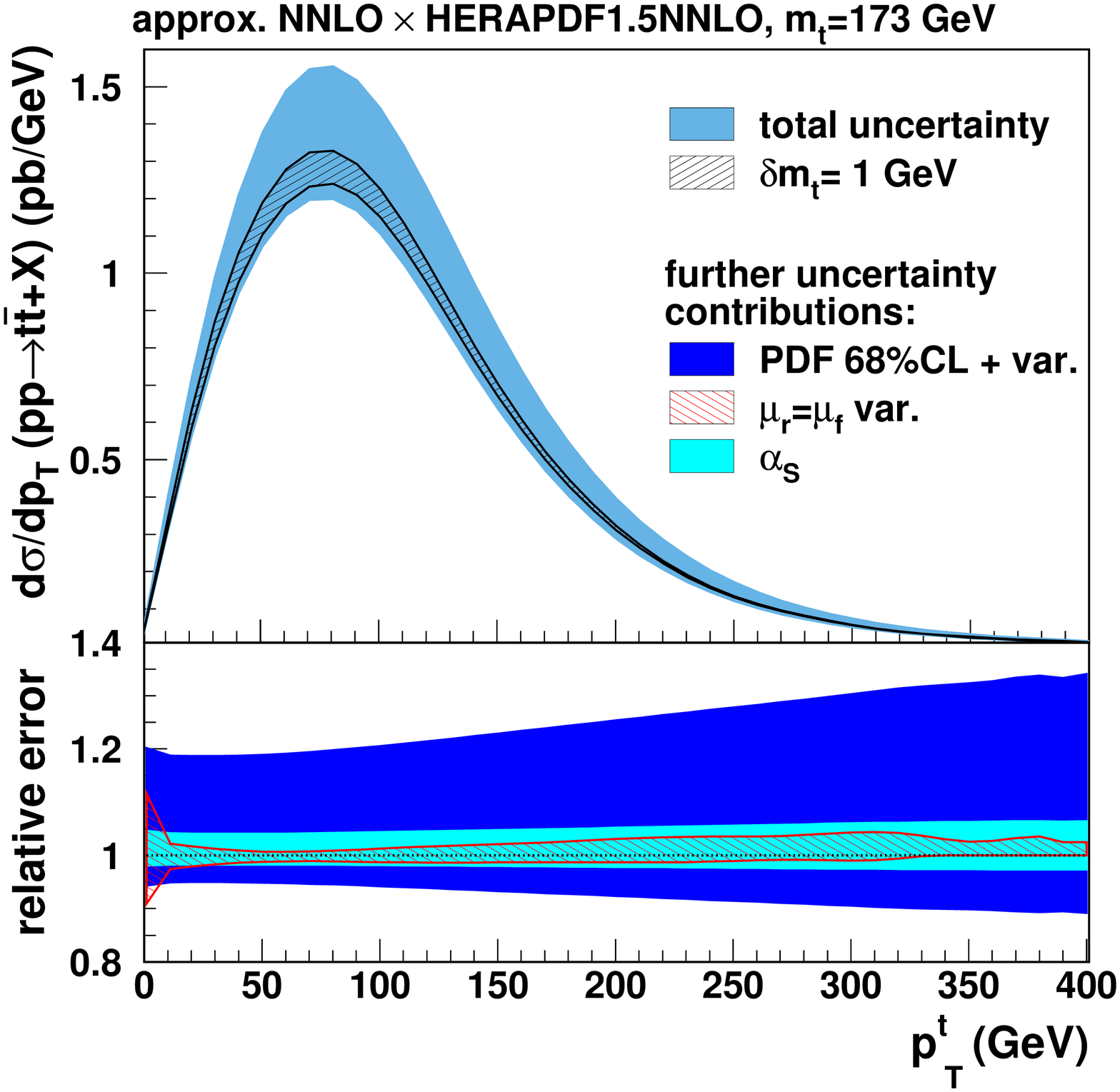}
\includegraphics[width=5.5cm, angle=0]{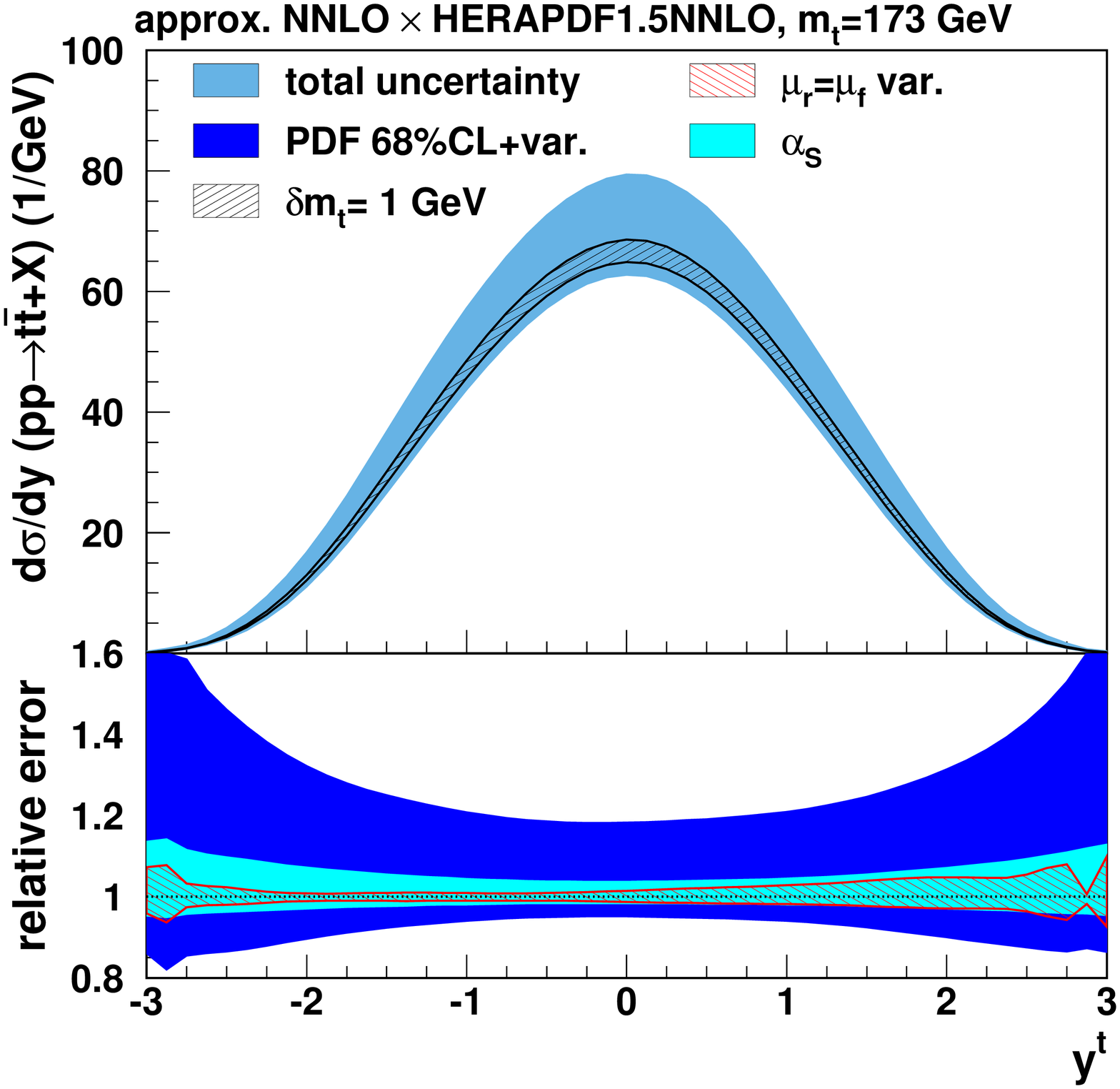}
\\
\includegraphics[width=5.5cm, angle=0]{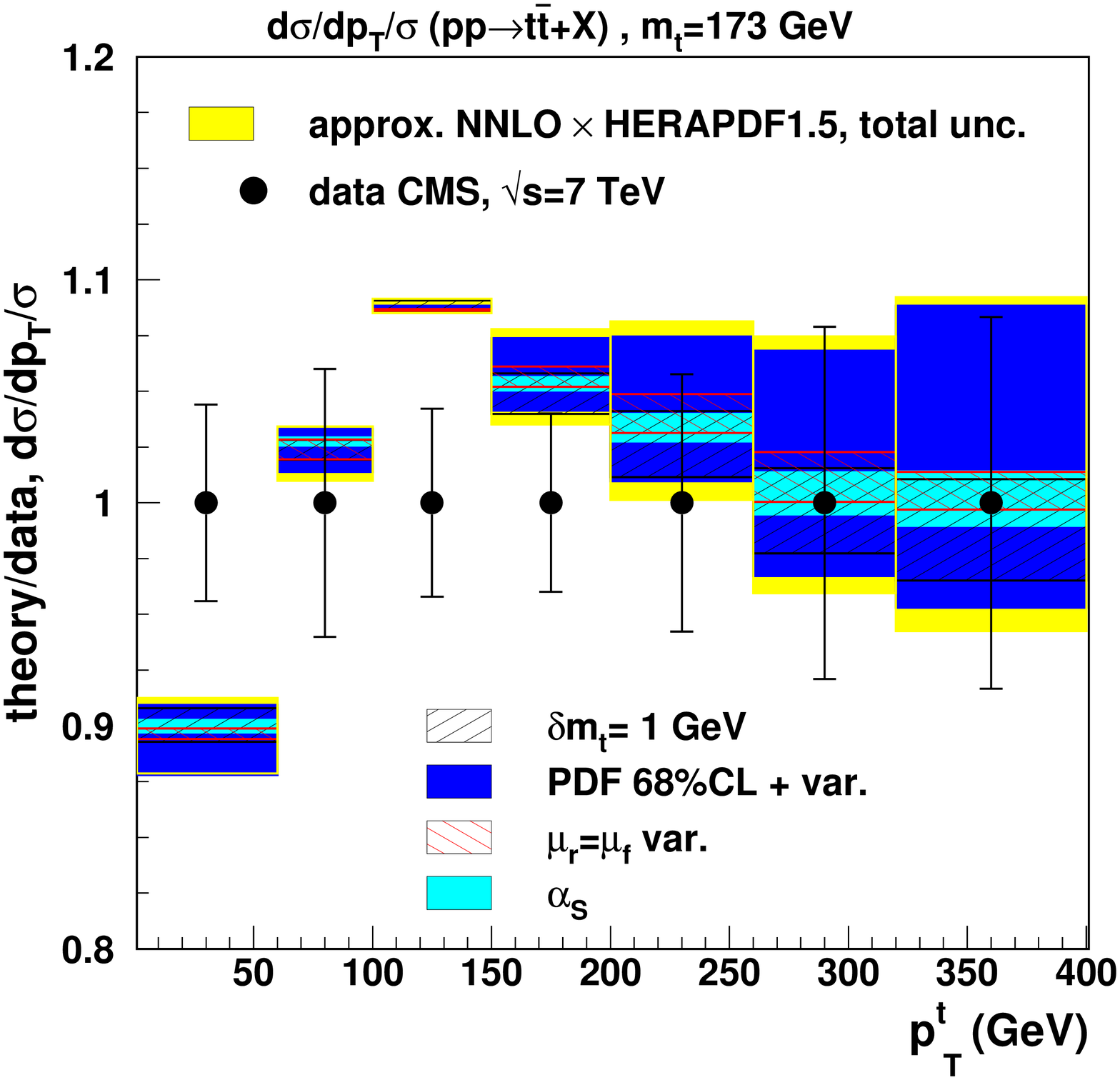}
\includegraphics[width=5.5cm, angle=0]{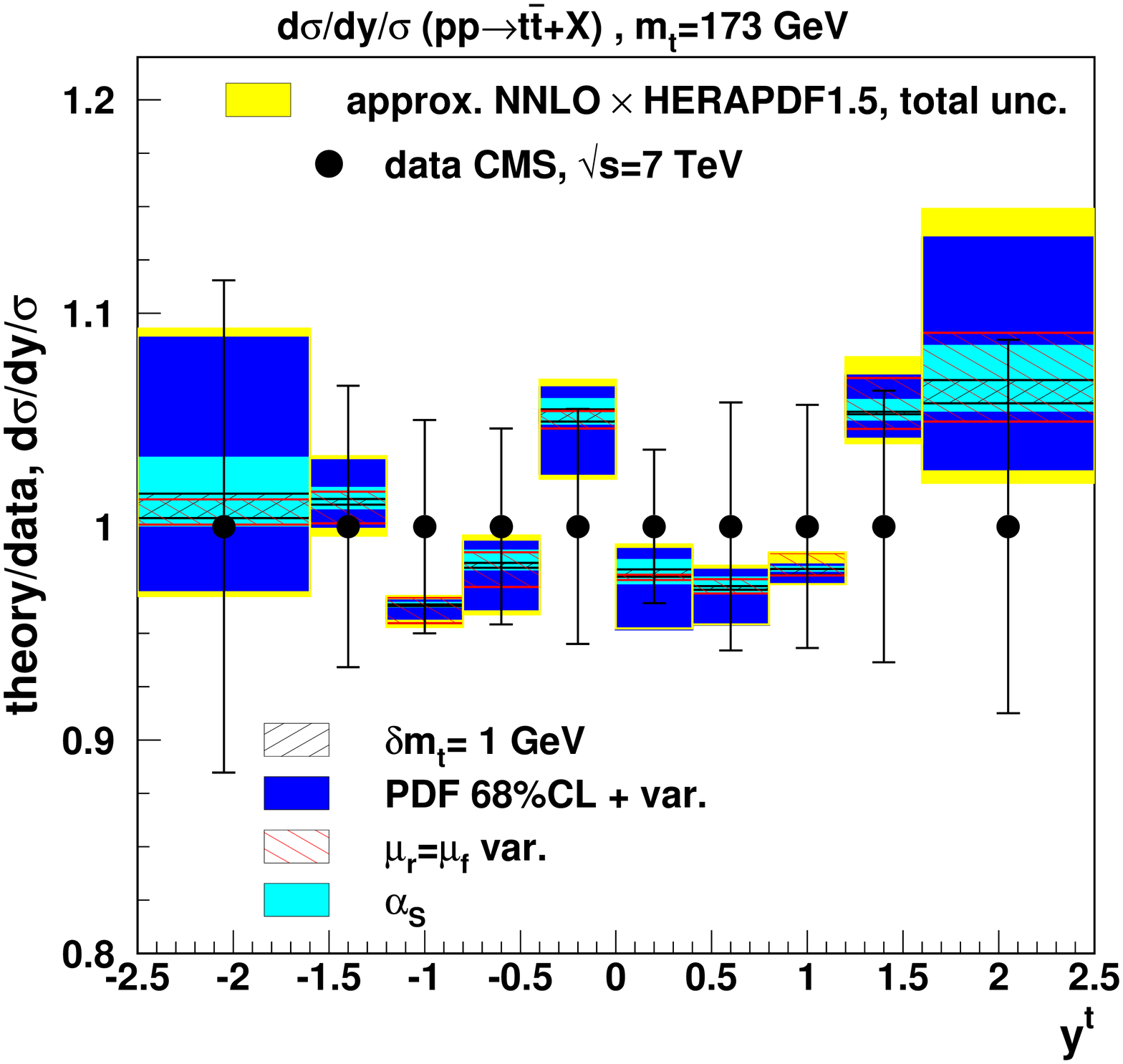}
\\
\includegraphics[width=5.5cm, angle=0]{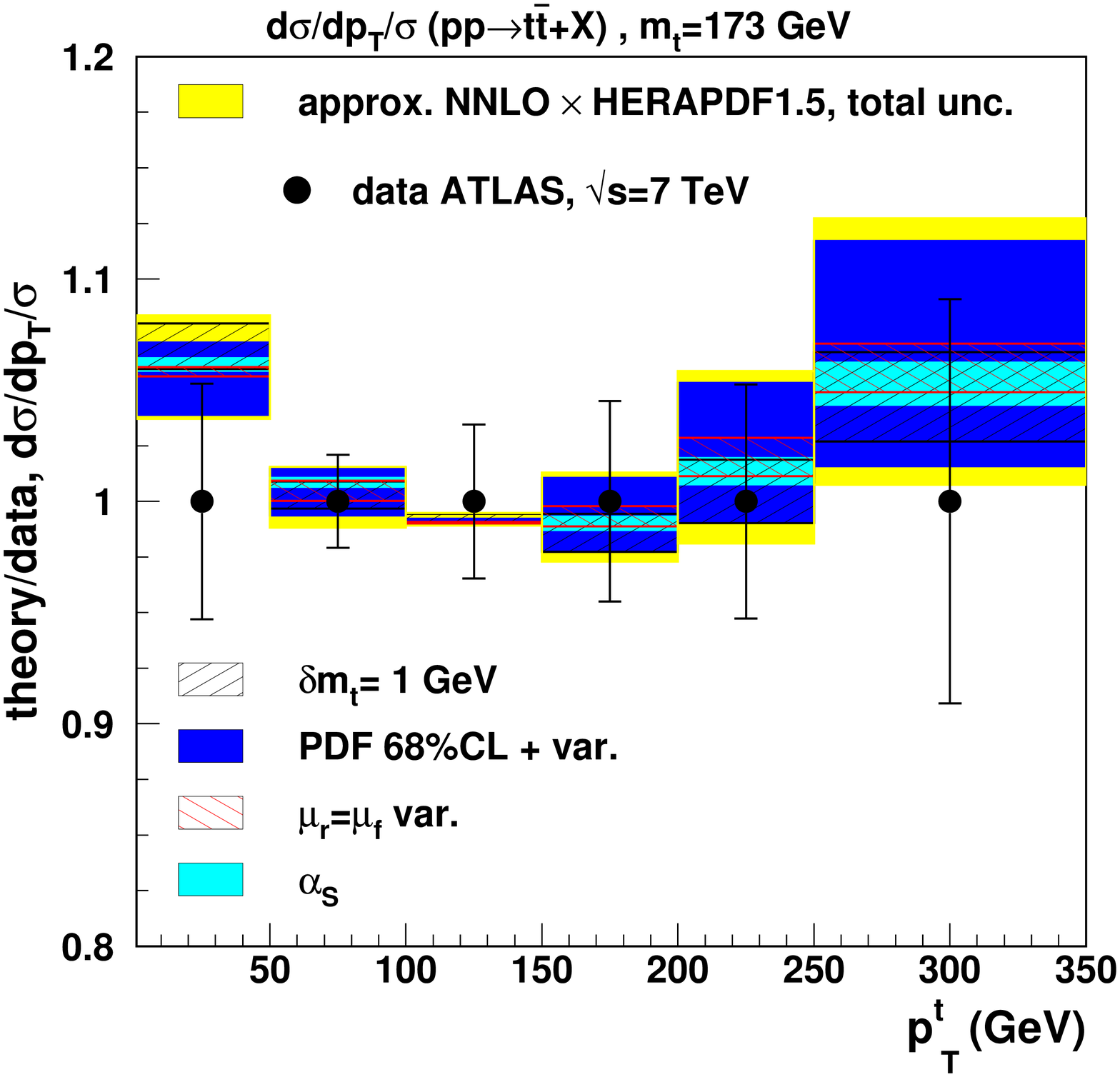}
\end{center}
\caption{Same as in Fig.~\ref{mstw08-unc} using HERAPDF1.5 NNLO. The experimental uncertainties are given at 68\% CL. 
In addition, model and parametrization uncertainties are considered.
\label{hera-unc}}
\end{figure}
\begin{figure}
\begin{center}
\includegraphics[width=5.5cm, angle=0]{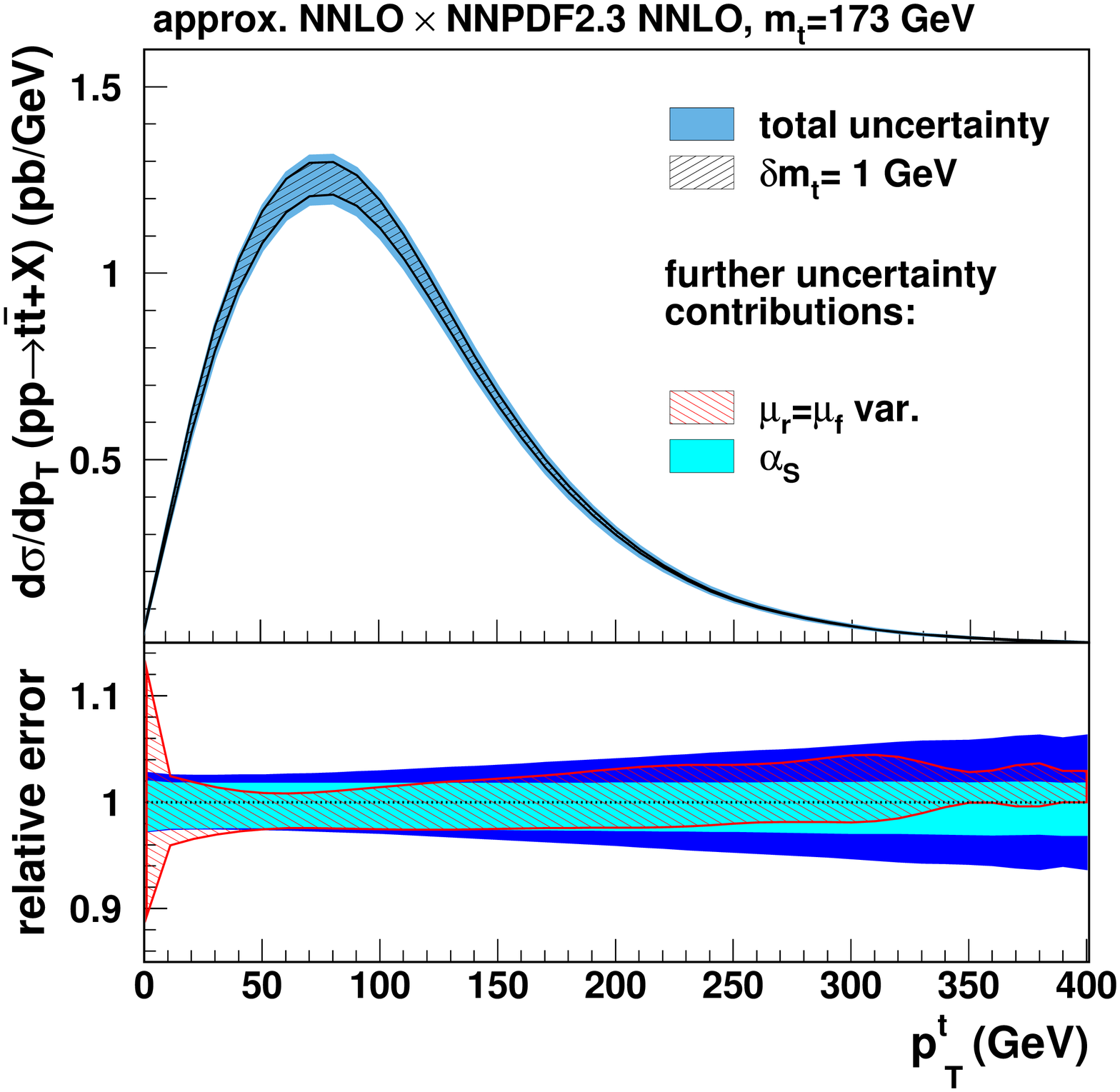}
\includegraphics[width=5.5cm, angle=0]{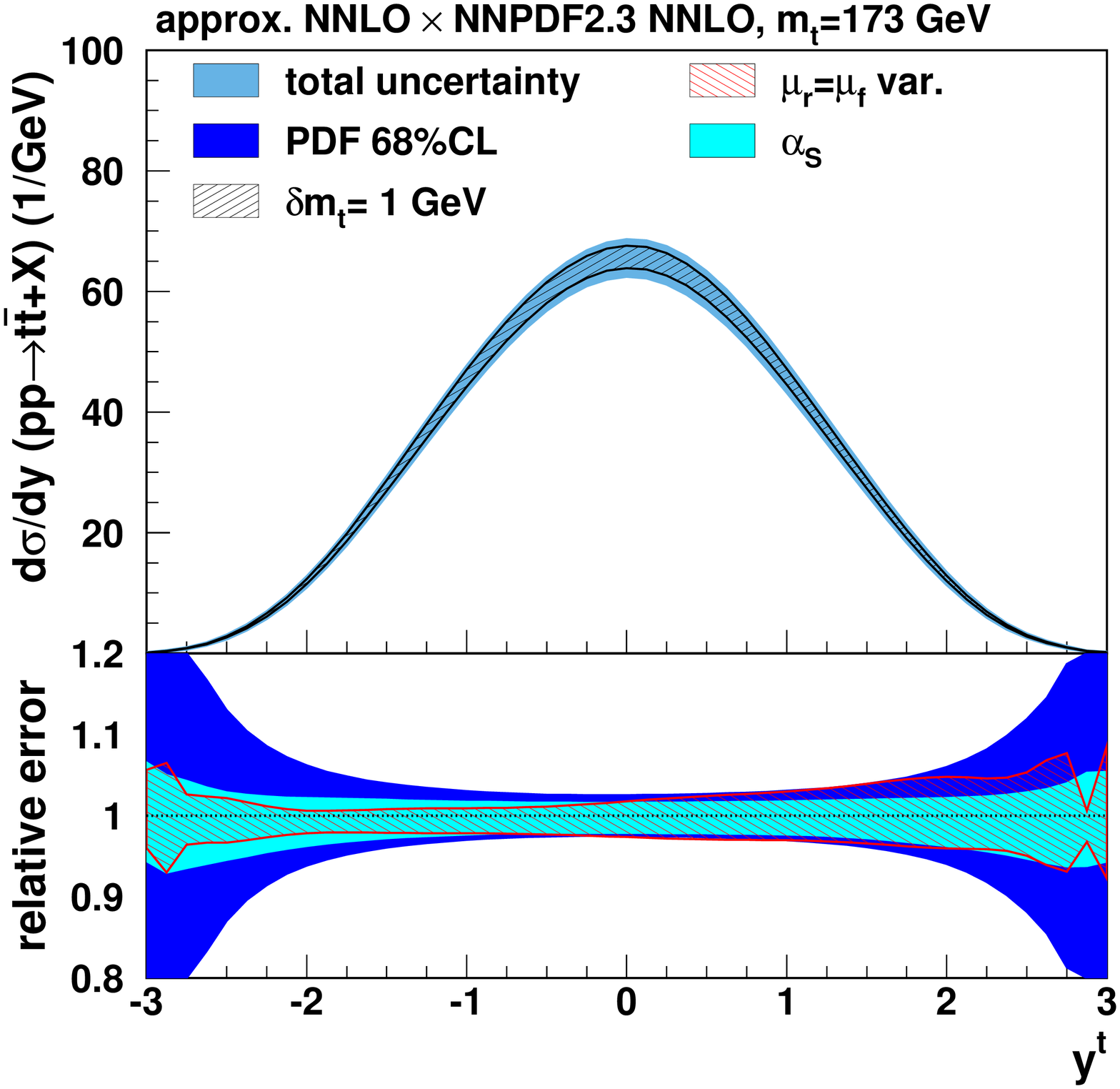}
\\
\includegraphics[width=5.5cm, angle=0]{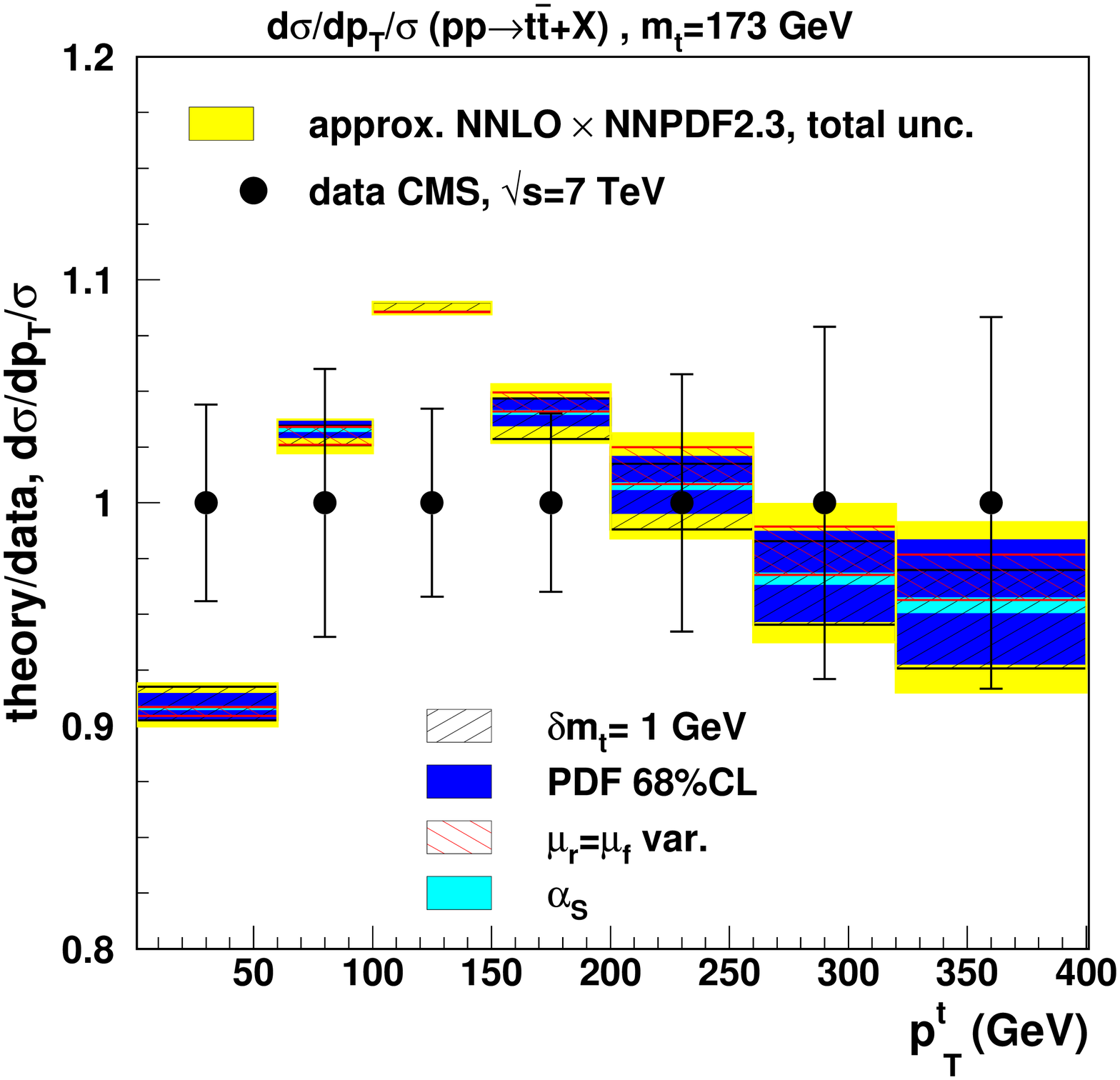}
\includegraphics[width=5.5cm, angle=0]{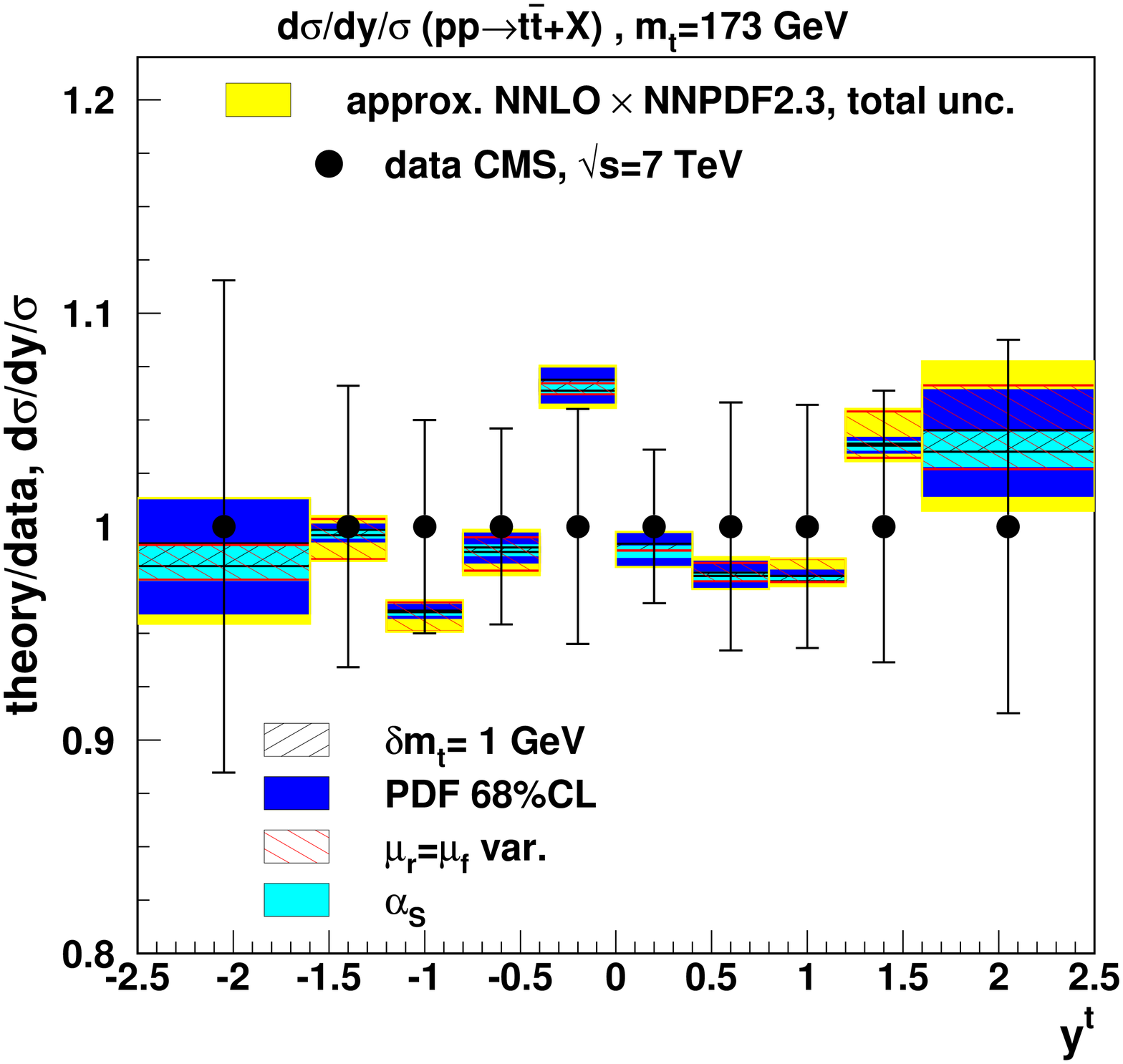}
\\
\includegraphics[width=5.5cm, angle=0]{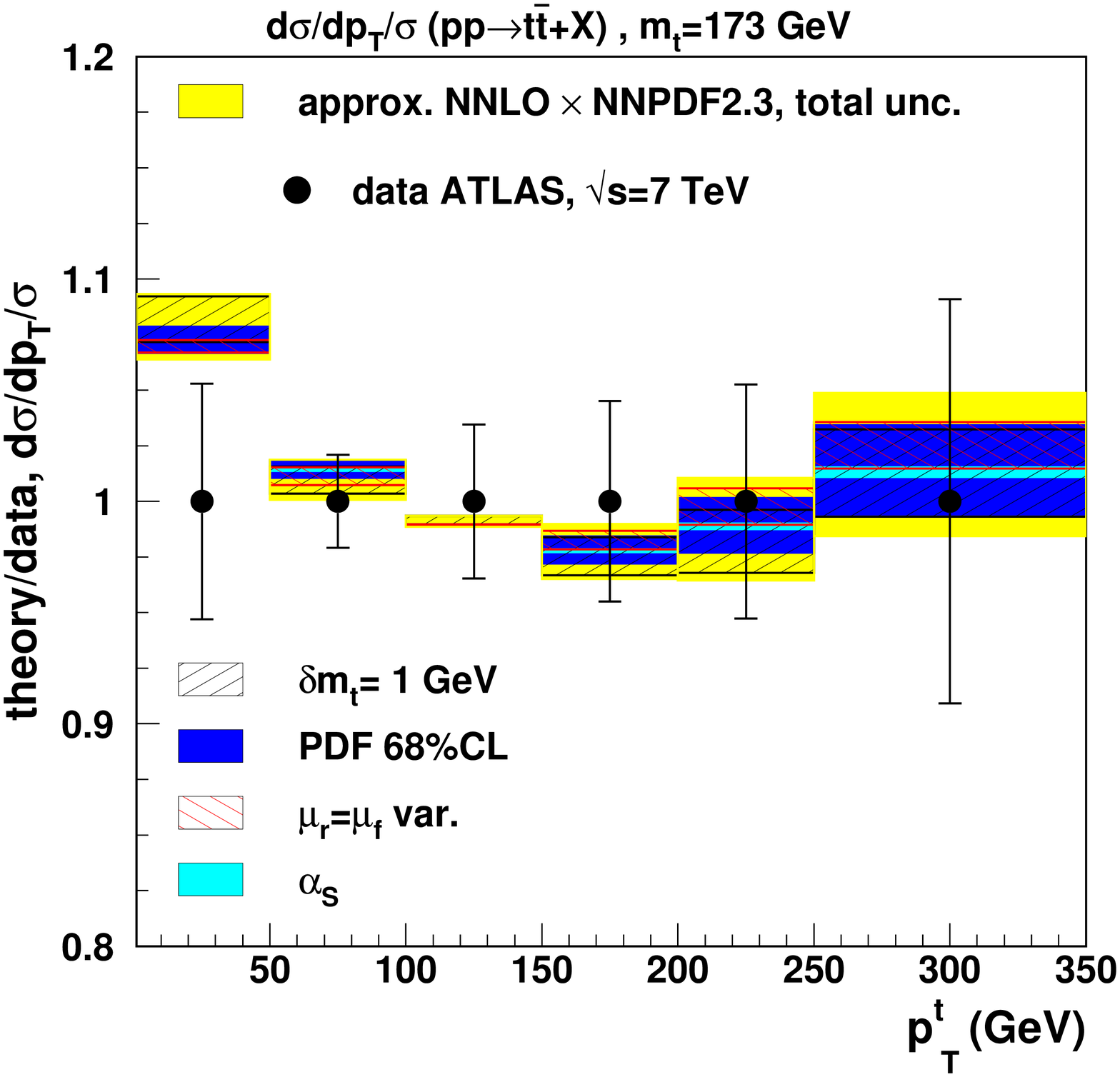}
\end{center}
\caption{Same as in Fig.~\ref{mstw08-unc} using NNPDF2.3 PDFs.
\label{nnpdf23-unc}}
\end{figure}

\end{document}